\documentclass[aps,prd,twocolumn,nofootinbib,longbibliography,10pt]{revtex4-2}
\usepackage{bm}
\usepackage{latexsym}
\usepackage{subfig}
\usepackage{amsmath,amsfonts,amssymb}
\usepackage{graphicx,epsfig}
\usepackage{psfrag}
\usepackage{amsthm}
\usepackage{bigints}
\usepackage{xcolor}
\usepackage{amsmath}
\usepackage[colorlinks=true,urlcolor=blue,citecolor=red,linkcolor=blue]{hyperref}
\usepackage{braket}
\usepackage{indent first}
\usepackage[normalem]{ulem}
\usepackage{multirow}
\useunder{\uline}{\ul}{}
 \usepackage{environ}
\usepackage{mathtools}
\usepackage{float}
\usepackage{relsize}

\DeclarePairedDelimiter\abs{\lvert}{\rvert}
\def\o{\omega}

\def\a{\alpha}
\def\b{\beta}
\def\k{\kappa}

\def\G{\Gamma}
\def\d{\delta}
\def\D{\Delta}

\def\t{\tau}
 
\def\p{\partial}
\def\f{\frac}

\def\tnudl{\widetilde{\nu_{dl}}}

\newcommand{\be}{\begin{equation}}
\newcommand{\ee}{\end{equation}}
\newcommand{\bes}{\begin{equation*}}
\newcommand{\ees}{\end{equation*}}

\newcommand{\beq}{\begin{eqnarray}}
\newcommand{\eeq}{\end{eqnarray}}

\NewEnviron{eqn}{
\begin{align}
\begin{split}
  \BODY
\end{split}
\end{align}
}

\NewEnviron{eqn*}{
\begin{align*}
\begin{split}
  \BODY
\end{split}
\end{align*}
}

\begin{document}
%\twocolumn[\begin{@twocolumnfalse}
\title{\textbf{Horizon brightened accelerated radiation in the background of braneworld black holes }}
\author{\textbf{Ashmita Das}}
\email{ashmita.d@srmap.edu.in}
\affiliation{Department of Physics, School of Engineering and Sciences, SRM University AP, Amaravati 522240, India, \& S.N. Bose National Centre for Basic Sciences, Department of Astrophysics and High Energy Physics, JD Block, Sector III, Salt Lake, Kolkata 700106, India}
\author{\textbf{Soham Sen${}^{a\dagger}$ and Sunandan Gangopadhyay${}^{a\ddagger}$}}
\affiliation{{${}^a$ Department of Astrophysics and High Energy Physics,}\\
{S.N. Bose National Centre for Basic Sciences,}\\
{JD Block, Sector III, Salt Lake, Kolkata 700106, India}}
%\date{}
%\maketitle
\begin{abstract}
%%%
\noindent
The concept of horizon brightened acceleration radiation (HBAR) has brought to us a distinct mechanism of particle production in curved spacetime.
In this manuscript we examine the HBAR phenomena for a braneworld black hole (BBH) which emerges as an effective theory in our $(3+1)$ dimensional universe due to the higher dimensional gravitational effects. Despite being somewhat similar to the Reissner-Nordstr$\ddot{\rm o}$m solution in general relativity, the BBH is unique with respect to its  charge term which is rather the tidal charge. 
%This tidal charge is not equal to the elctrical charge of the black hole. 
In this background, we study the transition probability of the atom due to the atom-field interaction and the associated HBAR entropy. Both the quantities acquire modifications  over the standard Schwarzschild results and turn out to be the function of the tidal charge. This modifications appear solely due to the bulk gravitational effects as induced on the 3-brane. 
%%Later, we compare the wavelengths associated to the maximum transition probability of the atoms as obtained for Schwarzschild and the BBH by implementing the Wien's displacement.
%%
Studying the Wien's displacement, we observe an important feature that the wavelengths of HBAR corresponding to the Schwarzschild and the BBH, deviate from each other depending on their masses. 
This deviation is found to be more pronounced 
for the mass values slightly greater or comparable to the Planck mass.
\end{abstract}

\maketitle

%\end{@twocolumnfalse}]
%
%
%
\section{Introduction}\label{intro}
\noindent\let\thefootnote\relax\footnote{{}\\
{$*$ashmita.phy@gmail.com}\\
{$\dagger$sensohomhary@gmail.com, soham.sen@bose.res.in}\\
{$\ddagger$sunandan.gangopadhyay@gmail.com}} 
In recent times some observational setup have been hypothesised to detect the thermal radiation due to the black holes and Unruh-Fulling (UF) effect \cite{Scully2003zz, Fulling2}. 
These are designed in order to probe the curved/flat spacetime by studying the interaction of atomic detector and quantum fields. Since the work by Scully \textit{et. al.} \cite{Scully2003zz}, this alternative mechanism, which implements the techniques from quantum optics to study the acceleration radiation in curved/flat spacetime, earned significant attraction. It was shown in \cite{Scully2003zz} that the transition probability for an accelerated atomic detector, interacting with single mode of photon within a high quality factor microwave cavity can be significantly increased than that of the standard transition probability in UF effect. 
%%
%It is well known that in quantum theory of fields, the virtual processes due to the vacuum fluctuations have disclosed many new quan- tum phenomena such as Lamb shift in the hydrogen atom which led to the construction of quantum electrodynamics, Raman scattering in the field of spectroscopy etc. 
Their work is conceptualised by the virtual processes which we frequently encounter in quantum field theory such as Lamb shift, Raman scattering, etc. In terms of atom-field interaction, a two level atom makes a transition to its excited state with the simultaneous emission of a virtual photon. Subsequently, the atom promptly comes back to its ground state by absorbing the emitted photon. It was shown that the UF effect can be perceived by interrupting a virtual process as a result of which the emitted virtual photon turns into the real photon \cite{Scully2003zz}. 
Later, this alternative technique has been implemented in case of a black hole spacetime where the two levels atomic detectors are freely falling in the black hole spacetime while passing through a cavity \cite{Fulling2}. The cavity is placed near the event horizon of the black hole in order to restrict the exposure of the detector from the Hawking radiated particles. A mode selector is designed which selects one cavity mode traversing in the direction opposite to that of the infalling atoms. This generates a relative acceleration between the atom and the field mode. This relative acceleration is the sole ingredient for the occurrence of acceleration radiation in the black hole spacetime.    
In particular due to the acceleration the atom gets away from its original point of virtual emission and triggers a nonzero probability of not absorbing the emitted photon. This transforms a virtual photon into a real one in the final state of the system \cite{Scully2003zz,Fulling2}. Subsequently, these radiated photons get detected by an asymptotic observer and the transition probability for the atoms along with the temperature and the entropy associated with this radiation can be examined. This kind of radiation which originates out of a different mechanism than the Hawking radiation, is known as horizon brightened acceleration radiation (HBAR) and the associated entropy is named as the HBAR entropy. 
It is worth to be mentioned that the atom acquires acceleration by extracting energy from some external force agency which drives the centre of mass motion of the atom \cite{Scully2003zz,Fulling2}. This mechanism has revealed many interesting findings which can be found in the following references \cite{Fulling2,Fulling,Ordonez1,Ordonez2,Ordonez3,Ordonez4,OTM, OTM2, Das:2022qpx}.
Note that until now all the literature on HBAR are developed in the background of  four dimensional black hole spacetime emerging from Einstein's general relativity (GR).  
Therefore it is an immediate question to ask that what would be the fate of the HBAR phenomena in the context of the alternative theories of gravity. Undoubtedly, extra dimensional theories are a few of the very important frameworks in the genre of alternative theories of gravity. Studying this HBAR phenomena in the context of the extra dimensional black hole spacetime would be a new entrant.

\noindent Within the domain of gravitational Physics extra dimensional models are competent to explain some of the observational phenomena such as the late time accelerated expansion of the universe, galaxy rotation curve, which otherwise successful Einstein's theory of GR cannot fully decode \cite{Clifton:2011jh, SupernovaCosmologyProject:1998vns, SupernovaSearchTeam:1998fmf}.  In general,  extra dimensional models  are described in  $D>4$ dimensional spacetime, where $D$ is the dimension of the higher dimensional spacetime and our $(3+1)$ dimensional Universe emerges as a hypersurface of this bulk higher dimensional spacetime, known as the visible/ TeV brane. Writing the bulk gravitational field equations, the $(3+1)$ dimensional effective Einstein's equation can be derived on the visible brane while projecting the bulk quantities on to the brane. Among the large variety of extra dimensional theories the noncompact geometry of extra dimension is of particular interest and these models are significant in exploring the cosmological aspects. One of these kind of theories is the 5-dimensional  Randall-Sundrum (RS) warped geometry model, abbreviated as RS2 \cite{Randall:1999vf}. From the functional perspective this model consists of single positive tension brane resembling our universe, while the negative tension brane is located at an infinite distance, resulting in a noncompact large extra dimensional scenario. For a literature survey on the RS2 model and some of its implications, we refer our readers to the following literature \cite{Maartens:2003tw,Langlois:2002bb,Brax:2004xh, Rubakov:2001kp}. In the background of RS2 model, Dadhich \textit{et. al.} have shown that an exact static and spherically symmetric black hole solution can be obtained on the visible brane which almost coincides with the Reissner-Nordstr$\ddot{\rm o}$m (RN) solution\cite{Dadhich:2000am}. However, the charge of the $(3+1)$ dimensional effective black hole spacetime is not electric charge by nature, rather it is tidal charge originating from the bulk curvature effect. The appearance of the tidal charge in $(3+1)$ dimensional effective theory is solely due to the gravitational effects in five spacetime dimensions, where such tidal correction term emerges in addition to the Schwarzschild potential. In this manuscript we name this black hole as BBH. 
Remarkably, this tidal charge appears in the linear order of the BBH metric and thus its negative value can be considered as an independent case. Rather it is a new possibility which we never encounter in the RN solution of GR. Therefore this black hole spacetime on the visible brane brings out two distinctive features than the black hole solutions in GR such as : 1) it resembles with the RN black hole solution with a gravitational charge instead of an electric charge of the black hole and 2) the gravitational charge could be a negative charge which is never possible for a standard RN metric. Several literature based on the BBH can be found in \cite{Emparan:2008eg,Maartens:2010ar,Kanti:2004nr,Hossenfelder:2003dy,Zhou:2011vy,Harmark:2004rm}.

These interesting features of BBH has led us to explore the phenomena of HBAR in this background. We consider the BBH solution as obtained in \cite{Dadhich:2000am} and study the transition probability of the atoms due to the atom-field interaction and the corresponding HBAR entropy emerging due to acceleration radiation of the atom.
We summarise our findings as follows
\begin{enumerate}
\item
We compute the transition probability while allowing upto the quadratic order of the tidal charge. This leads to the modified transition probability where the modifications become proportional to the quadratic power of the tidal charge. This outcome signifies that the transition rate is indifferent to the sign of the tidal charge. 
\item
HBAR entropy exhibits a similar area-entropy relation as one obtains for the standard black hole solutions. However it acquires modifications which are dependent on the tidal charge of the BBH. 
\item
Implementing the theory of Wien displacement, we present a comparative study between the wavelengths of the emitted radiation which correspond to the maximum transition probability in both the standard Schwarzschild and BBH spacetime. 
The wavelength corresponding to the standard Schwarzschild spacetime depends on the parameters such as the mass of the black hole $(M)$ and the 4 dimensional Planck mass ($M_p$). Whereas the same for the BBH turns out to be the function of $M, \,M_p,\,q,\,\tilde{M}_p$, where $q, \tilde{M}_p$ denote the tidal charge and five dimensional Planck mass respectively. 
\item
The tidal charge ($q$) can be constrained by Eq.   (\ref{radius_2}). We obey this condition throughout the phenomenological exploration of the model.
\item
Exploring the parameter space of the wavelengths, we get certain amount of deviations in their values, notably in the region where the mass of the black hole ($M$) is slightly greater or equal to the Planck mass $(M_p)$. 
\item
We also examine the variation of the wavelengths with respect to the tidal charge where the plot for standard Schwarzschild black hole shows no alteration. However, a decreasing pattern in the wavelength of the emitted radiation with increasing tidal charge can be noted for the BBH. 
\end{enumerate}
These alterations in the HBAR radiation spectrum, entropy and incoming wavelength of the emitted radiation are due to the effective description of the bulk gravitational degrees of freedom on the $(3+1)$ dimensional brane. For the first time, our study reveals the fate of HBAR radiation in the background of an extra dimensional theory. 
\subsection{Black hole solution on the brane}\label{BBH_1}
\noindent In extra dimensional scenario the bulk gravitational equations are considered to be 
higher dimensional Einstein's field equations. Therefore, the lower dimensional effective theory can be extracted from the higher dimensional Einstein's equations while taking the projections of the bulk equations and parameters on to the brane. 
In this section, we briefly discuss the construction of a  black hole solution on the visible brane due to the higher dimensional curvature effect as described in \cite{Dadhich:2000am}. The authors of \cite{Dadhich:2000am} have implemented a ``\textit{geometrical approach}" to perceive the five dimensional RS2 model, the technique of which has been introduced by Shiromizu \textit{et. al.} in \cite{Shiromizu:1999wj}. 
The projection of the bulk metric $\widetilde{g}_{AB}$ on the brane turns out to be $g_{AB}=\,\widetilde{g}_{AB}\, - n_A n_B$, 
where $g_{AB}$ depict the induced brane metric and 
$n_A$ is the normal to the brane. 
According to the proposition in \cite{Shiromizu:1999wj}, upon using the bulk gravitational equation, the Gauss-Codazzi equation which yields the projection of five  dimensional Riemann curvature tensor on the visible brane and $Z_2$ symmetry of the extra dimension, one obtains the effective gravitational equations on the brane. The branes are located at the orbifold fixed points and the visible brane (our universe) is at $\chi =0$, where $\chi$ symbolises the extra dimensional coordinate. Therefore, the $(3+1)$-dimensional gravitational field equation becomes \cite{Dadhich:2000am},  
\beq
G_{\mu\nu}= - \Lambda g_{\mu\nu}+ \k^2 T_{\mu\nu}+\tilde{\k}^4 S_{\mu\nu }-\mathcal{E}_{\mu\nu}
\label{eins_4d_1}
\eeq
where $G_{\mu\nu}$, $g_{\mu\nu}$ denote the effective Einstein tensor and metric on the visible brane and $\lambda$ denotes the brane tension. In the above equation $\k^2=\f{8 \pi}{M_{p}^{2}}$ and $\tilde{\k}^2=\f{8\pi}{\tilde{M}_{p}^{3}}$. 
%where $\tilde{M}_{p},\,M_p$ are the five and four dimensional Planck mass.
 Here, $\Lambda$ is the induced cosmological constant on the brane. These set of bulk and brane parameters are associated with each other as below,
\beq
M_p=\sqrt{\f{3}{4 \pi}} \left(\f{\tilde{M}_{p}^{3}}{\sqrt{\lambda}}\right),\,\,\,\,\,\,\,\,\Lambda=\f{4 \pi}{\tilde{M}_{p}^{3}}\left[\tilde{\Lambda}+\left(\f{4 \pi}{3 \tilde{M}_{p}^{3}}\right) \lambda^2\right]~.
\eeq
In the above equation $\tilde{\Lambda}$ denotes the bulk cosmological constant. In the RS2 scenario, $\tilde{M}_p$ can be considered to be much smaller than $M_p$.  
$T_{\mu\nu}$ is the net energy-momentum tensor on the brane and $S_{\mu\nu}$ is the squared energy-momentum tensor which carries the effect of the bulk on the matter fields residing on the visible brane \cite{Dadhich:2000am}. For the present purpose, we intend to derive the bulk solution of the gravitational equations which demands $S_{\mu\nu}=T_{\mu\nu}=0$. Moreover considering $\tilde{\Lambda}=-\f{4 \pi \lambda^2}{3 \tilde{M}_{p}^{3}}$, one can produce zero cosmological constant on the brane, that is $\Lambda =0$. 
Thus, Eq.   (\ref{eins_4d_1}) becomes, 
\beq
R_{\mu\nu}= - \mathcal{E}_{\mu\nu},\,\,\,\,\,\,\,\,\,R= 0=  \mathcal{E}_{\mu}\,^{\mu}
\label{eins_4d_2}
\eeq
where $R_{\mu\nu}$ is the induced Ricci tensor and $\mathcal{E}_{\mu\nu}$ is the projection of the bulk Weyl tensor on the visible brane. This $\mathcal{E}_{\mu\nu}$ can be written as, 
\beq
\mathcal{E}_{\mu\nu}= ^{(5)}C_{\mu a \nu b}\, n^a n^b~.
\label{weyl_1}
\eeq
Eq.   (\ref{eins_4d_2}) dictates that the induced Weyl curvature term can be portrayed as the source term on the brane representing the effects of nonlocal gravitational degrees of freedom in the bulk. Exploiting Weyl symmetry one can write that 
$\mathcal{E}_{\mu\nu}$ is symmetric and traceless. For a vacuum solution on the visible brane one gets,
\beq
\nabla^{\mu} \mathcal{E}_{\mu\nu} = 0, 
\label{bianchi-1}
\eeq
 due to the Binachi identity, where $\nabla^{\mu}$ is the covariant derivative defined with respect to the metric on the visible brane. Equation $\nabla^{\mu} \mathcal{E}_{\mu\nu} = 0$ and Eq.   (\ref{eins_4d_2}) form a set of closed equations which upon solving yields the geometry of the visible brane.  For a detailed discussion, we refer our readers to \cite{Dadhich:2000am,Maartens:2000fg}. 

For the solutions of these equations, it was shown that the spherically symmetric and static solution can be achieved on the visible brane by decomposing $\mathcal{E}_{\mu\nu}$ in the form of irreducible representation in terms of four velocity $u_{\mu}$. 
This produces two equations such as for the effective energy density on the brane ($\mathcal{U} (r)$) and the anisotropic stress ($\mathcal{P}(r)$). Here $r$ symbolises the radial distance. Note that $\mathcal{E}_{\mu\nu}$
is antisymmetric and tracefree. Thus, it exhibits similar algebraic properties as the energy momentum tensor of the electromagnetic field \cite{Dadhich:2000am,Whisker:2008kk}. This implies that the effective bulk Weyl term on the visible brane has a correspondence with the electromagnetic energy-momentum tensor in GR, that is 
$-\mathcal{E}_{\mu\nu}  \leftrightarrow T_{\mu\nu}^{(em)}$. 
Similarly the conservation equation for the energy momentum tensor in GR corresponds to Eq.   (\ref{bianchi-1}) in the context of braneworld scenario. 
This has led the authors of \cite{Dadhich:2000am} to consider a $(3+1)$-dimensional effective theory on the visible brane where a tidal correction term resembling the RN correction term, appears along with the Schwarzschild potential in the metric. This correction term can be depicted as, 
\beq
\Phi =\, - \f{M}{M_{p}^{2} r}+ \f{Q}{ 2 r^2}~.Eq.   
\label{pot_1}
\eeq
Choosing the equation of state as  $\mathcal{U}+ \f{\mathcal{P}}{2} = 0$, one 
obtains the solution for conservation equation (Eq.   (\ref{bianchi-1})) as, $\mathcal{U} = \left(\f{\k}{\tilde{\k}}\right)^4\, \f{Q}{r^4}$. This solution is compatible to Eq.   (\ref{pot_1}) and it can be confirmed that Eq.  (\ref{pot_1}) and the solution for $\mathcal{U} $ satisfy Eq.  (\ref{eins_4d_2}). The spacetime metric corresponding to these solutions turns out to be as follows, 
%
%The induced black hole spacetime on the $(3+1)$ dimensional brane:
%
%
\beq
ds^2= -f(r) dt^2+ \f{dr^2}{f(r)}+r^2\,(d\theta^2+\sin^2 \theta\,d\phi^2)
\label{metric_1}
\eeq
where, $f(r)= 1-\f{2 M }{M_{p}^{2}\,r}+\f{Q}{r^2}$. 
The charge $Q$ can be redefined in terms of a dimensionless charge $q$ as $q= Q \tilde{M}_{p}^2$. Thus, the lapse function $f(r)$ in terms of the new dimensionless charge $q$ can be redefined as $f(r) = 1-\f{2 M }{M_{p}^{2}\,r}+\f{q}{\tilde{M}_{p}^2 r^2}$. 
Below we briefly mention the properties of such a BBH. 
\begin{itemize}
\item
Note that in Eq.  (\ref{metric_1}), the tidal charge correction term is linear in $q$ which is different than that of the RN solution in GR.  Therefore, the features of this BBH will depend on the sign of $q$. For the radius of the horizon, the $(3+1)$-dimensional standard RN solutions emerge when $q \geq 0$ as below,
\beq
r_{\pm}=\f{M}{M_{p}^{2}}\left[1 \pm \sqrt{1-\f{q \,M_{p}^{4}}{M^2\, \tilde{M}_{p}^{2}}}\right]
\label{radius_1}~.
\eeq   
Similar to the RN black hole in GR, these two horizons fall within the Schwarzschild radius $r_s= \f{2 M}{M_{p}^{2}}$, that is 
$r_{-} \leq r_{+} \leq r_s$. 
\item
Some distinctive features can be perceived for the case $q < 0$. This predicts the existence of one horizon, however, located outside $r_s$. This can be depicted as follows
\beq
r_{+}= \f{M}{M_{p}^{2}}\left[1 + \sqrt{1-\f{q \,M_{p}^{4}}{M^2\, \tilde{M}_{p}^{2}}}\right]
\label{radius_2}~.
\eeq
In GR we never encounter such possibilities. This enlarged radius of the horizon implies a larger Bekenstein-Hawking entropy and reduction in the temperature of the BBH than the Schwarzschild case. Therefore, $q<0$ case suggests that  the bulk gravitational effects assist to create a stronger gravitational field on the visible brane.  On the other hand as the $q>0$ case for the BBH exactly matches with the RN black hole in GR, this suppresses the strength of the gravitational field on the visible brane. A more detailed discussion can be found in \cite{Dadhich:2000am,Whisker:2008kk} (sections (4-5)).  
\end{itemize} 
\subsection{Trajectory of a freely falling atomic detector in the BBH}
\noindent In this section, we study the trajectory of a freely falling atomic detector in the background of a BBH with the metric ansatz given in Eq.   (\ref{metric_1}). The set of equations which yields the trajectory of the detector are as follows, 
\begin{subequations}
     \begin{equation}
        \f{dt}{dr}= -\,\f{1}{f(r)\,\sqrt{1-f(r)}},\,\,\,\,\,\,\,\,\,\,\,\,\,\,
                     \,\,\,\,\f{dt}{d\t}= \f{1}{f(r)}~,
                     \label{traj_1}
     \end{equation}
     \begin{equation}
       \,\,\,\,\,
            \,\,\,\,\,\,\,\,\f{d\t}{dr}=\,- \f{1}{\sqrt{1-f(r)}}~.
            \label{traj_2}
      \end{equation}
\end{subequations}
%\begin{subequations}
%     \begin{equation}
%        \f{dt}{dr}= -\,\f{1}{f(r)\,\sqrt{1-f(r)}},\,\,\,\,\,\,\,\,\,\,\,\,\,\,
%                     \,\,\,\,\f{dt}{d\t}= \f{1}{f(r)}
%                     \label{traj_1}
%     \end{equation}
%     \begin{equation}
%        \,\,\,\,\,\,\,\,\,\,\,\,\,\,\,\f{dr_{*}}{dr}=\,\f{1}{f(r)},\,\,\,\,\,\,\,\,\,\,\,\,\,\,\,\,\,\,\,\,
%            \,\,\,\,\,\,\,\,\f{d\t}{dr}=\,- \f{1}{\sqrt{1-f(r)}}
%            \label{traj_2}
%      \end{equation}
%\end{subequations}
%
%
%
We take the induced tidal charge $Q$ to be a small quantity and allow upto its quadratic order ($\mathcal{O}(Q^2)$) in our analysis. Expanding $\frac{1}{f(r)}$ upto the quadratic order in $Q$, we obtain, 
\begin{subequations}
     \begin{equation}
\f{1}{f(r)} \approx \bigg(1-\f{2 M}{M_{p}^{2} r}\bigg)^{-1}\,\bigg[1-\f{Q}{r \left(r-\f{2 M}{M_{p}^{2}}\right)}+ \f{Q^2}{r^2 \left(r-\f{2 M}{M_{p}^{2}}\right)^2}\bigg]\label{fr_1}
\end{equation}
\begin{equation}
\f{1}{\sqrt{1-f(r)}} \approx\, M_{p}\,\sqrt{\f{r}{2 M}}\,\left[ 1+\, \f{Q \,M_{p}^{2}}{4\,r\,M} +\, \f{3 \,Q^2\, M_{p}^{4}}{32\, M^2\, r^2}\right]~.
\label{fr_2}
\end{equation}
\end{subequations}
Integrating Eq.  (s)(\ref{traj_1},\ref{traj_2}), one gets, 
\beq
t(r)&=&-\,\f{2 \sqrt{2}\, M^{1/2} \,r^{1/2}}{M_p} - \f{M_p}{3}\,\sqrt{\f{2}{M}}\,r^{3/2}\nonumber\\
&&+\,\f{2 M}{M_{p}^{2}} \,{\rm ln}\,\abs*{\f{1+\,M_p\,\sqrt{\f{r}{2M}}}{1-\,M_p\,\sqrt{\f{r}{2M}}}}+\,\f{Q\,M_{p}^{5}\,r^{3/2}}{(2 M)^{\frac{3}{2}}\,(2 M- M_{p}^{2}\, r)}\nonumber\\
&&+\,\f{7\sqrt{2}\,Q^2\,M_{p}^{3}}{16\,(r\,M)^{3/2}}
\left(1-\f{2 M}{M_{p}^{2}\,r}\right)^{-2}\,\left[1-\f{5 M_{p}^{2}\,r}{14\,M}\right]\nonumber\\
&&+\,\f{M_{p}^{6}\,Q^2}{8\,M^3}\,{\rm ln}\,\abs*{\f{1+\,M_p\,\sqrt{\f{r}{2M}}}{1-\,M_p\,\sqrt{\f{r}{2M}}}}+\,{\rm const.}
\label{traj_t}
\eeq
\beq
\t(r)=\,-\,\f{M_p\,r^{3/2}}{3}\,\sqrt{\f{2}{M}}-\,\f{Q\,M_{p}^{3}\,r^{1/2}}{(2 \,M)^{3/2}}&+&\,\f{3\,Q^2\,M_{p}^{5}\,r^{-1/2}}{16\,\sqrt{2}\,M^{5/2}}\nonumber\\
&+&\,{\rm const.}~.
\label{traj_tau}
\eeq
Defining, 
 \begin{equation}
\f{dr_{*}}{dr}=\,\f{1}{f(r)}, 
\end{equation}
we obtain, 
\beq
r_{*}(r)&=&\,r+\,\f{2 M}{M_{p}^{2}}\,{\rm ln}\,\abs*{\f{M_{p}^{2}\,r}{2\,M}-\,1}+\,Q\,\left(r-\f{2\,M}{M_{p}^{2}\,r}\right)^{-1}\nonumber\\
&&+\,\f{M_{p}^{6}\,Q^2}{8\,M^3}\,{\rm ln}\,\abs*{1-\f{2\,M}{M_{p}^{2}\,r}}+\,\f{M_{p}^{6}\,Q^2}{4\,M^2}\,\f{M_{p}^{2}\,r-\,3\,M}{(M_{p}^{2}\,r-\,2\,M)^2}\nonumber\\
&&+\,{\rm const.}~.
\label{traj_rstar}
\eeq
These set of equations is depicting the trajectory of the freely falling atomic detector as a function of radial distance $(r)$ in the BBH spacetime. 
At this stage we introduce the dimensionless forms of the parameters $(r,\,\t,\, t,\, Q,\, \o,\,\nu)$ using $r_{+}$ as the unit of distance \cite{Fulling2}. We consider $c=1$, and write down the parameters as follows, 
\begin{subequations}
\begin{equation}
\,\,r=\,r_{+}\,r_{dl}~,\,\,\,\,\,\,\,\,\,\,\,\,\,\,\,\,\,\,\,\,\, t=\,r_{+}\,t_{dl}~,\,\,\,\,\,\,\,\,\,\,\,\,\,\,\,\,\,\,\,\,\, \t=\,\,r_{+}\,\t_{dl}~,
\label{diml_1}
\\
\end{equation}
\begin{equation}
Q=\,r_{+}^{2}\,Q_{dl}\,\,\,\,\,\,\,\,\,\,\,\,\,\, \o=\,\f{\o_{dl}}{r_{+}}\,\,\,\,\,\,\,\,\,\,\,\,\,\,\, \nu=\,\f{\nu_{dl}}{r_{+}}\,\,\,\,\,\,\,\,\,\,
\,g^{2}_{dl}=\,g^2\,r_{+}^{2}~.
\label{diml_2_f}
\end{equation}
\end{subequations}
Here all the parameters with the subscript $dl$ symbolise the dimensionless form. Using this parametrisation, Eq.  (s)(\ref{traj_t}), (\ref{traj_tau},\ref{traj_rstar}) become,
\beq
t_{dl}(r_{dl})&=&- 2 r_{dl}^{1/2}- \f{2 r_{dl}^{3/2}}{3}+(1+Q_{dl}+Q_{dl}^{2})\nonumber\\
&&{\rm ln} \abs*{\f{\sqrt{r_{dl}}+1}{\sqrt{r_{dl}}-\,1}}-\,2 Q_{dl} \sqrt{r_{dl}}+\,\f{Q_{dl}\,r_{dl}^{3/2}}{3}\nonumber\\
&&-\,\f{Q_{dl}^{2}\,r_{dl}^{1/2}\,(r_{dl}-\,1)}{4}+\,\f{Q_{dl}^{2}\,r_{dl}^{1/2}}{4\,(r_{dl}-\,1)}\nonumber\\
&&-\,\f{7\,Q_{dl}^{2}\,r_{dl}^{1/2}}{4\,(r_{dl}-\,1)}
+\,\f{3\,Q_{dl}^{2}\,r_{dl}^{3/2}}{2\,(r_{dl}-\,1)}
\label{diml_t_1}
\eeq
\beq
\t_{dl}(r_{dl})&=&\,-\,\f{2\,r_{dl}^{3/2}}{3}+\,Q_{dl}\,\left(\f{r_{dl}^{3/2}}{3}-\,r_{dl}^{1/2}\right)\nonumber\\
&&+\,Q_{dl}^{2}\,\left(\f{3\,r_{dl}^{1/2}}{2}+\,\f{3\,r_{dl}^{-1/2}}{4}-\,\f{r_{dl}^{3/2}}{4} \right)~,
\label{diml_tau_1}
\eeq
\beq
r_{*}^{dl} (r_{dl})&=&\,r_{dl}+\,(1+Q_{dl}+Q_{dl}^{2})\,{\rm ln}\,\abs*{r_{dl} -\,1}-\,Q_{dl}\nonumber\\
&&-\,\f{Q^{2}_{dl}}{2}-\,Q^{2}_{dl}\,{\rm ln}\,r_{dl}~.
\label{diml_r_1}
\eeq
We further obtain, 
\beq
(r_{*}^{dl}-t_{dl}) (r_{dl})&=&r_{dl}+2 r_{dl}^{1/2}+ \f{2 r_{dl}^{3/2}}{3}\nonumber\\
&&+2 (1+Q_{dl}+Q_{dl}^{2}) {\rm ln}\abs*{\sqrt{r_{dl}} -1}\nonumber\\
&&-Q_{dl}\left(1-2 r_{dl}^{1/2}+ \f{r_{dl}^{3/2}}{3}\right) \nonumber\\
&&-\f{Q_{dl}^{2}}{2}\bigg(1+ 2 {\rm ln} r_{dl}+\f{7 \sqrt{r_{dl}}}{2}- \f{r_{dl}^{3/2}}{2}\bigg)~.\nonumber\\
\label{diml_r_star_1}
\eeq
These equations describe the trajectory of the atomic detector as a function of $r_{dl}$ in the background of the BBH. 
\section{Acceleration radiation from the freely falling atoms in the BBH spacetime}
In this section we follow the procedure developed in \cite{Scully2003zz,Fulling2} and consider the Klein Gordon equation for a massless scalar photon with wave function $\Psi$ as, 
\beq
\f{1}{\sqrt{-g}}\p_{\mu}\left(\sqrt{-g}\,g^{\mu\nu} \p_{\nu}\right)\Psi = 0~.
\eeq
Imposing the s-wave approximation in the above equation one obtains, 
\beq
\f{1}{T(t)}\f{d^2 T}{d t^2} - \f{f(r)}{r^2 R(r)} \f{d}{d r}\left(r^2 f(r)\f{d R(r)}{d r}\right) = 0,
\label{scalar_eqn_1}
\eeq
where using the method of separation of variables we write, $\Psi (t, r)=T(t)\,R(r)$. Subsequently, one can write the general solution for the Eq.  (\ref{scalar_eqn_1}) as follows, 
\beq
\Psi_{\nu} (t, r)= {\rm exp}\left[- i \nu t + i \nu \int\, \f{dr}{f(r)}\right].
\label{scalar_eqn_2}
\eeq
Here, $\nu$ depicts the frequency of the photon field as detected by the asymptotic observer. 
We now turn our attention to examine the transition probability of the freely falling detector while interacting with the field mode. 

\noindent The interaction Hamiltonian for the system reads 
\begin{equation}
\hat{\mathcal{H}}_I(\tau)=\hbar \mathcal{G}\left[\hat{b}_{\nu}\Psi_{\nu}+h.c.\right]\,\left[\hat{\sigma}e^{-i\omega\tau}+h.c.\right], 
\label{hamiltonian_int}
\end{equation}
which leads one to obtain the transition probability as follows, 
\beq
P_{\rm exc}=\,g^2\,\abs*{\int d\t\,\,e^{i\nu t(r)-i \nu r_{*}(r)}\,e^{i\o \t (r)}}^2.
\label{prob_1}
\eeq
%
%
%%
%\beq
%P_{\rm exc}=\,g^2\,r_{+}^{2}\abs*{\int d\t_{dl}\,e^{-i\nu_{dl} (r_{*}^{dl} -\,t_{dl})}\,e^{i\o_{dl} \t_{dl}}}^2
%\label{prob_1a}
%\eeq
%%
%
We recast Eq.  (\ref{prob_1}) in terms of the dimensionless parameters as follows,
\beq
P_{\rm exc}=\,g^{2}_{dl}\,\abs*{\bigintsss_{r_{dl}=\infty}^{1}\,\,\,\f{d\t_{dl}}{d r_{dl}}\,d r _{dl}\,\,e^{-i\nu_{dl} (r_{*}^{dl}-\,t_{dl})}\,e^{i\o_{dl} \t_{dl}}}^2
\label{prob_2}, 
\eeq
where,
\beq
\f{d\t_{dl}}{dr_{dl}}&=&\,-\,r_{dl}^{1/2}+\,\f{Q_{dl}}{2}\,\left[r_{dl}^{1/2}-\,r_{dl}^{-1/2}\right]\nonumber\\
&&+\,\f{3\, Q_{dl}^{2}}{8}\,\bigg[2\,r_{dl}^{-1/2}-\,r_{dl}^{-3/2}-\,r_{dl}^{1/2}\bigg]. 
\eeq
Changing the variable $r_{dl}^{3/2}=\,y$ in the above equations we obtain,
\beq
\t_{dl}(y)&=&\,- \f{2\,y}{3}+\,Q_{dl}\,\bigg(\f{y}{3}-\,y^{1/3}\bigg) \nonumber\\
&&+\,Q_{dl}^{2}\,\bigg(\f{3\,y^{1/3}}{2} +\,\f{3\,y^{-1/3}}{4} -\,\f{y}{4}\bigg)~,
\eeq 
\beq
(r_{*}^{dl}- t_{dl}) (y)&=& y^{2/3} +2 {\rm ln}\abs*{y^{1/3}-1} +2 y^{1/3}+\f{2 y}{3}\nonumber\\
&-&Q_{dl} \bigg(1-2 {\rm ln}\abs*{y^{1/3}-1} -2 y^{1/3}+\f{y}{3}\bigg) \nonumber\\
&-&\f{Q_{dl}^{2}}{2}\bigg(1-4{\rm ln}\abs*{1-y^{-1/3}} +\f{7 y^{1/3}}{2} -\f{y}{2}\bigg)\nonumber~,
\\
\label{diml_r_star_2}
\eeq
\beq
\f{d \t_{dl}}{d r_{dl}}(y)&=&ty^{1/3}+\f{Q_{dl}}{2}\bigg(y^{1/3}-y^{-1/3}\bigg)\nonumber\\
&&+\f{3 Q_{dl}^{2}}{8}\bigg(2 y^{-1/3}-y^{-1}- y^{1/3}\bigg)~.
\label{deriv_tau_diml_1}
\eeq
We write the transition probability in terms of $y$ in Appendix (\ref{app_1}). 

To perform the integration, we change the integration variable $y=\,1+\f{3\,x}{2\,\o_{dl}}$, where we take $\o_{dl}\gg\,1$, and keep upto $\mathcal{O}\left(\f{x}{\o_{dl}}\right)^2$ in subsequent analysis. Note that  within the logarithmic terms we keep upto the cubic order of the same. 
Thus, the above equations turn out to be 
\beq
\t_{dl}(x)=\,- \f{2}{3}-\,\f{x}{\o_{dl}}+\,Q_{dl}\,\bigg(\f{x^2}{4\,\o_{dl}^{2}} -\,\f{2}{3}\bigg)+\,2\,Q_{dl}^{2}~,
\label{tau_diml_x_1}
\eeq
\beq
(r_{*}^{dl}-t_{dl}) (x)&=&\f{11}{3}+\f{2 x}{\o_{dl}}-\f{x^2}{6 \o_{dl}^{2}}+ 2 {\rm ln} \Big(\f{x}{2 \o_{dl}}\Big)\nonumber\\
&&- \f{Q_{dl}}{2}\bigg[\f{x}{\o_{dl}}-\f{x^2}{6 \o_{dl}^{2}}- 4 {\rm ln} \Big(\f{x}{2\o_{dl}} \Big)-\f{4}{3}\bigg]\nonumber\\
&-& 2Q_{dl}^{2} \bigg[1- {\rm ln}\Big(\f{x}{2\o_{dl}} \Big)+\f{5 x}{4 \o_{dl}}- \f{85 x^2}{96 \o_{dl}^{2}}\bigg]\nonumber~,\\
\eeq
\beq
\f{d \t_{dl}}{d r_{dl}}(x)&=&- \bigg(1+\f{x}{2 \o_{dl}}- \f{x^2}{4 \o_{dl}^{2}}\bigg) + \f{Q_{dl}}{2} \bigg(\f{x}{\o_{dl}}\nonumber\\
&&- \f{3 x^2}{4 \o_{dl}^{2}} \bigg)- \f{3 Q_{dl}^{2}}{8} \f{x^2}{\o_{dl}^{2}}~.
\label{deriv_tau_diml_2}
\eeq
In terms of these new parametrisation, we get the excitation probability as follows, 
\beq
P_{\rm exc}&=& \f{g^{2}_{dl}}{\o_{dl}^{2}}\Bigg|\bigintsss_{0}^{\infty}d x \left[1-\f{Q_{dl} x}{2 \o_{dl}}+\f{5 Q_{dl} x^2}{8 \o^{2}_{dl}}+\f{3 Q_{dl}^{2} x^2}{8 \o^{2}_{dl}}\right]\nonumber\\
&&e^{-i\nu_{dl} \phi(x)}\,{\rm Exp}\Bigg[i\,\o_{dl}\Bigg\{-\f{2}{3}- \f{x}{\o_{dl}}+ \f{Q_{dl} x^2}{4 \o_{dl}^{2}}- \f{2 Q_{dl}}{3}\nonumber\\
&&+ 2 Q_{dl}^{2}\Bigg\}\Bigg]
\Bigg|^2~.
\label{prob_2_f}
\eeq
Here, 
\beq
\phi (x)&=&\, \f{11}{3}+ \f{2 \,x}{\o_{dl}}- \f{x^2}{6\, \o_{dl}^{2}}+ 2\, {\rm ln}\,\f{x}{2\, \o_{dl}}\nonumber\\
&&+ Q_{dl}\bigg[2\,{\rm ln}\,\f{x}{2\, \o_{dl}}- \f{x}{2\, \o_{dl}}+ \f{2}{3}+ \f{x^2}{12\,\o_{dl}^{2}}\bigg]\nonumber\\
&&+ Q_{dl}^{2} \bigg[ 2\,{\rm ln}\,\f{x}{2\, \o_{dl}}- \f{5\, x}{2\, \o_{dl}}- 2+ \f{85\,x^2}{48\,\o_{dl}^{2}}\bigg]~.\nonumber\\
\label{phi_1}
\eeq
Upon simplification one gets, 
\beq
P_{exc}&=&\f{g^{2}_{dl}}{\o_{dl}^{2}}\Bigg|\bigintssss_{0}^{\infty} d x \left[1-\f{Q_{dl} x}{2 \o_{dl}}+\f{5 Q_{dl} x^2}{8 \o^{2}_{dl}}+\f{3 Q_{dl}^{2} x^2}{8 \o^{2}_{dl}}\right]\nonumber\\
&&\bigg[e^{- 2 i \tnudl  {\rm ln} x- i x+ \f{i Q_{dl}x^2}{4 \o_{dl}}}\bigg] {\rm Exp}\bigg[- \f{2 i \nu_{dl} x}{\o_{dl}}\bigg(1-\f{Q_{dl}}{4}\nonumber\\
&&- \f{5 Q_{dl}^{2}}{4}\bigg)
+ \f{i \nu_{dl} x^2}{6 \o_{dl}^{2}}\left(1- \f{Q_{dl}}{2}- \f{85 Q_{dl}^{2}}{8}\right)\bigg]\Bigg|^{2}
\label{prob_3}
\eeq
where, $\tnudl =\, \nu_{dl}\,\left(1+Q_{dl}+ Q_{dl}^{2}\right)$.
One can also write Eq.  (\ref{prob_3}) as follows, 
\beq
P_{exc}&=&\,\f{g^{2}_{dl}}{\o_{dl}^{2}}\,\,\Bigg|\bigintssss_{0}^{\infty}\,d x\, \left( 1- A\,x+ B\,x^2 \right)\,x^{-2 i\,\tnudl}\nonumber\\
&&e^{-i p_1\,x+\,i p_2\,x^2}\Bigg|^2,
\label{prob_5}
\eeq
where,
\begin{subequations}
\begin{equation}
\,\,A=\f{Q_{dl}}{2\,\o_{dl}},\,\,\,\,\,\,\,\,\,\,\,\,\,\,\,\,\,\,\,\,\,\,\,B= \f{(5+ 3\,Q_{dl})\,Q_{dl}}{8\,\o_{dl}^{2}},
\label{diml_1_f}
\\
\end{equation}
\begin{eqnarray}
p_1=1+ \f{2\,\nu_{dl}}{\o_{dl}}\left( 1- \f{Q_{dl}}{4}- \f{5\,Q_{dl}^{2}}{4}\right)\,\,\,\,{\rm and}\\
p_2=\,\f{Q_{dl}}{4\,\o_{dl}} + \f{\nu_{dl}}{6\,\o_{dl}^{2}}\left(1- \f{Q_{dl}}{2} - \f{85\,Q_{dl}^{2}}{8}\right)~.
\label{diml_2}
\end{eqnarray}
\end{subequations}
The exact integration of Eq.  (\ref{prob_5}) yields hypergeometric function which is quite  complex to tackle. Thus, we approximate $e^{i p_2\,x^2} \sim\, (1+ i p_2\,x^2)$, where it is considered that $p_2\,x^2\,<\,1$ for any value of $x$. 

\noindent 
This approximation is legitimate as $\o_{dl}^{2} \gg\,1$ appears in the denominator of $p_2$ while a small parameter such as $Q_{dl}$ appears in the numerator.  Therefore, their ratio turns out to be a small quantity and one can proceed with the leading order approximation.
Therefore, the transition probability becomes, 
\beq
P_{exc}&=& \f{g^{2}_{dl}}{\o_{dl}^{2}} \Bigg|\bigintssss_{0}^{\infty} d x \left( 1- A x+ B x^2 \right) (1+ i p_2 x^2)\nonumber\\
&& x^{-2 i \tnudl} e^{-i p_1 x}\Bigg|^2 = \f{g^{2}_{dl}}{\o_{dl}^{2}} \Big|\,I\,\Big|^2~.
\label{prob_6}
\eeq
We compute the above integral in Appendix (\ref{int_1}) which leads us to the form of the transition probability as follows
\beq
P_{exc}&=& \f{4\,\pi g^{2}_{dl} \,\nu_{dl}}{\o_{dl}^{2}\left(1+\f{2 \nu_{dl}}{\o_{dl}}\right)^2}\bigg[ 1+ Q_{dl} + 3\, Q_{dl}^{2} \left(1 + \f{4 \,\nu_{dl}}{3\, \o_{dl}}\right)\bigg]\nonumber\\
&&\times\f{1}{e^{4 \pi\,\nu_{dl}(1+ Q_{dl}+ Q_{dl}^{2})} - 1}~.
\label{prob_7}
\eeq
At this stage we transform all the dimensionless parameters to their respective dimensionful counterpart and write the transition probability as below, 
\beq
P_{exc}= \f{4\,\pi g^{2} \,\nu\,r_{+}}{\o^{2}\left(1+\f{2 \nu}{\o}\right)^2}\,\f{1+\f{Q}{r_{+}^{2}}+\left(3+\f{4 \nu}{\o}\right)\f{Q^{2}}{r_{+}^{4}}}{e^{4 \pi\nu\,r_{+}\left(1+\f{Q}{r_{+}^{2}}+\f{Q^2}{r_{+}^{4}}\right)} -1}~.
\label{prob_8}
\eeq
In Eq.  (\ref{prob_8}) we use, 
\beq
r_{+}^{2}=\,\f{4\,M^2}{M_{p}^{4}}\left(1- \,\f{Q\,M_{p}^{4}}{4\,M^2} -\, \f{Q^2\,M_{p}^{8}}{16\,M^4}\right)^{2}
\label{r_plus_1}
\eeq
which leads to the following equations where we keep upto the quadratic order in $Q$.
\beq
&&\f{Q}{r_{+}^{2}}=\f{Q M_{p}^{4}}{4 M^2}\left(1- \f{Q M_{p}^{4}}{4 M^2} - \f{Q^2 M_{p}^{8}}{16 M^4}\right)^{-2}\nonumber\\
&&\approx \f{Q M_{p}^{4}}{4 M^2}+\f{Q^2 M_{p}^{8}}{8 M^4}~
\label{q_r_plus_1}
\eeq
\beq
{\rm and}\,\,\,\,\,\,\,\,\,\,\,\,\,\,\,\,\f{Q^2}{r_{+}^{4}} \approx \f{Q^2 M_{p}^{8}}{16 M^4}~.
\label{q_r_plus_2}
\eeq
Using these approximations, the excitation probability in Eq.   (\ref{prob_8}) takes the form 
\beq
P_{exc} &\approx& \f{4 \pi g^{2} \nu}{\o^{2}\left(1+\f{2 \nu}{\o}\right)^2}\Bigg[\f{2 M}{M_{p}^{2}}+\f{3 Q^{2} M_{p}^{6}}{8 M^3}\left(1+\f{4 \nu}{3\o}\right)\Bigg]\nonumber\\
&& \times\f{1}{e^{\f{8 \pi M \nu}{M_{p}^{2}}\left(1+\f{Q^2 M_{p}^{8}}{16 M^{4}}\right)} -1}~.
\label{prob_9}
\eeq
The final form of the transition probability depicts a dependence on the parameter $Q^{2}$ which is an induced gravitational charge due to the extra dimensional spacetime. 
A further simplification leads us to obtain, 
\beq
P_{exc} \approx \f{4\pi g^{2} \nu}{\o^{2}\left(1+\f{2 \nu}{\o}\right)^2}\Bigg[\f{2 M}{M_{p}^{2}}+ \f{3 Q^{2} M_{p}^{6}}{8 M^3} \Bigg]\nonumber\\
\f{1}{e^{\f{8 \pi M \nu}{M_{p}^{2}}\left(1+\f{Q^2 M_{p}^{8}}{16 M^{4}}\right)} -1}.
\eeq
It can be realised from the above equation that when $\nu \gg 1$ the transition probability becomes exponentially suppressed resulting no acceleration radiation. However, the atomic frequency $\o$ can be much greater than $1$. Therefore, for the occurrence of acceleration radiation it is legitimate to consider $\nu \ll \o$. This leads us to the transition probability as follows, 
\beq
P_{exc}\approx 
\f{4\,\pi g^{2} \,\nu}{\o^{2}}\,\Bigg[\f{2 M}{M_{p}^{2}}+\,\f{3\,Q^{2} M_{p}^{6}}{8\,M^3} \Bigg] \f{1}{e^{\f{8 \pi M \nu}{M_{p}^{2}}\left(1+\f{Q^2\,M_{p}^{8}}{16\,M^{4}}\right)} -1}\,.
\nonumber
\\
\label{prob_10}
\eeq
Proceeding similarly as above the absorption probability for the atom can be written as, 
\beq
P_{abs} &\approx& 
\f{4\,\pi g^{2} \,\nu}{\o^{2}}\,\Bigg[\f{2 M}{M_{p}^{2}}+\,\f{3\,Q^{2} M_{p}^{6}}{8\,M^3} \Bigg] \f{1}{1 - e^{-\f{8 \pi M \nu}{M_{p}^{2}}\left(1+\f{Q^2\,M_{p}^{8}}{16\,M^{4}}\right)}}\nonumber~.
\\
\label{prob_abs}
\eeq
\section{HBAR entropy of the BBH}
\noindent In this section, we aim to find the rate of change of entropy corresponding to the acceleration radiation in the background of the BBH spacetime. We follow the trick from quantum optics as used in \cite{Scully2003zz,Fulling2}, where obtaining the density matrix of the field is the first concern. Thus, we write the microscopic change in the field density matrix as, $\d \rho_{i}$ due to a single atom. Consequently, the macroscopic change in the same due to the $\D \mathcal{N}$ number of atoms can be written as \cite{Scully2003zz,Fulling2}, 
\beq
&&\Delta \rho=\, \sum_{i} \d \rho_i=\,\D \mathcal{N}\,\d \rho=\,\k\,\D t\,\d \rho
\label{den_fld_1}\\
&&\implies\, \f{\D \rho}{\D t}=\,\k\,\d \rho
\label{den_fld_2}
\eeq
Here $\k$ depicts the rate at which the atoms fall into the event horizon of the black hole. Using the Lindblad master equation for the density matrix one obtains, 
\beq
\dot{\rho}_{n,n}= &-& \G_{\rm abs} \left[n \rho_{n,n} - (n+1) \rho_{n+1, n+1}\right] \nonumber\\
&-& \G_{\rm exc} \left[(n+1) \rho_{n,n} - n \rho_{n-1, n-1}\right], 
\label{rho_lin}
\eeq
where, $\G_{\rm exc},\,\G_{\rm abs}$ symbolise emission and absorption rates of the photons in the cavity by the atom and these rates are defined as, $\G_{\rm exc/abs}= \k\,P_{\rm exc/abs}$. 

\noindent The steady state solution for the density matrix of the field becomes, 
\beq
\rho_{n,n}^{\mathcal{S}}=\, \left(\f{\G_{\rm exc}}{\G_{\rm abs}}\right)^{n}\,\left( 1- \f{\G_{\rm exc}}{\G_{\rm abs}}\right).
\label{rho_ss_1}
\eeq
This is the equation of motion for the density matrix of the emitted photon fields due to the HBAR. In \cite{Scully2003zz,Fulling2} and the present manuscript,  one can realise that there is a change in the mathematical treatment in this section. In the earlier sections the system is studied with respect to the acceleration radiation, transition probability of the atom, etc. However, in this section the mathematical treatment is performed in terms of the density matrix of the field $(\rho)$, its equation of motion, etc. Once the density matrix of the field is obtained, one can conveniently derive the Von Neumann entropy for the system. 

\noindent Due to the real photon production the time rate of change of entropy becomes,  
\beq
\dot{S}_\rho=\,-\,k_{B}\,\sum_{n,\nu}\,\dot{\rho}_{n,n}\,{\rm ln}(\rho_{n,n}).
\label{hbar_entropy_1}
\eeq
Using the steady state solution of the density matrix, the above equation can approximately be written as, 
\beq
\dot{S}_\rho\,\approx\,-\,k_{B}\,\sum_{n,\nu}\,\dot{\rho}_{n,n}\,{\rm ln}(\rho^{\mathcal{S}}_{n,n}).
\label{hbar_entropy_2}
\eeq
Furthermore using Eq.  (\ref{rho_ss_1}) in Eq.  (\ref{hbar_entropy_2}) we obtain, 
\beq
\dot{S}_\rho\,&=&\,-\,k_{B}\,\sum_{n,\nu}\,\dot{\rho}_{n,n} \Bigg[n\,{\rm ln}\,e^{-\,\f{8 \pi M \nu}{M_{p}^{2}}\left(1+\f{Q^2\,M_{p}^{8}}{16\,M^{4}}\right)}\nonumber\\
&&+\,{\rm ln} \left(1-\,e^{-\,\f{8 \pi M \nu}{M_{p}^{2}}\left(1+\f{Q^2\,M_{p}^{8}}{16\,M^{4}}\right)} \right)\Bigg]
\label{hbar_entropy_3}
\\
&\approx&\,k_{B}\,\sum_{n,\nu}\,n\,\dot{\rho}_{n,n} \,\left[\f{8 \pi M \nu}{M_{p}^{2}}\left(1+\f{Q^2\,M_{p}^{8}}{16\,M^{4}}\right)\right]\nonumber\\
&=&\,\f{8 \pi M k_B}{M_{p}^{2}}\left(1+\f{Q^2\,M_{p}^{8}}{16\,M^{4}}\right)\sum_{\nu}\dot{\bar{n}}_{\nu}\,\nu~.
\label{hbar_entropy_4}
\eeq
Here, $\dot{\bar{n}}_{\nu}$ depicts flux of the produced photons in the cavity. 
The area of the black hole can be written as $A_{\rm BH}=\,4 \pi\,r_{+}^{2}$. 
%Now one can write the change in the black hole mass as follows, 
%
%
%\beq
%\D M_{\rm BH}=\,\D M_{\rm photon}+\,\D M_{\rm atom}~.
%\label{mass_chnage_1}
%\eeq
%
%
One can now write the rate of change of the black hole mass as below,
%
%
%\beq
%\dot{\D M_{\rm BH}}=\,\D \dot{M}_{\rm photon}+\,\D \dot{M}_{\rm atom}~.
\beq
\dot{M}_{\rm BH}=\,\dot{M}_{\rm photon}+\,\dot{M}_{\rm atom}~.
\label{mass_change_time_1}
\eeq
The area of the black hole can be written as, 
\beq
A_{\rm BH}&=&4 \pi r_{+}^{2}= 4\pi\left(\f{2 M_{\rm BH}}{M_{p}^{2}}\right)^2\nonumber\\
&\times &\left(1-\f{Q\,M_{p}^{4}}{4\, M_{\rm BH}^{2}} - \f{Q^2\,M_{p}^{8}}{16\,  M_{\rm BH}^{4}}\right)^2\nonumber\\
&\approx& \f{16 \pi}{M_{p}^{4}}\,\left(M_{\rm BH}^{2}-\f{Q M_{p}^{4}}{2}-\f{Q^2 M_{p}^{8}}{16 M_{\rm BH}^{2}}\right)
\label{area_change_1}~.
\eeq
Thus, differentiating $A_{\rm BH}$ with respect to time we obtain, 
\beq
\dot{A}_{\rm BH}=\,\f{32 \pi\,M_{\rm BH}\,\dot{ M_{\rm BH}}}{M_{p}^{4}}\,\left(1+\f{Q^2\,M_{p}^{8}}{16\, M_{\rm BH}^{4}}\right)
\label{area_change_time_1}~.
\eeq
Further, we obtain, 
\beq
\f{\dot{A}_{\rm BH}}{A_{\rm BH}}&=& \f{2 \dot{M}_{\rm BH}}{M_{BH}} \left(1+\f{Q^2\,M_{p}^{8}}{16\,M_{\rm BH}^{4}}\right)\nonumber\\
&\times &\left(1-\f{Q\,M_{p}^{4}}{2\, M_{\rm BH}^{2}} - \f{Q^2\,M_{p}^{8}}{16\, M_{\rm BH}^{4}}\right)^{-1}\nonumber\\
&\approx & \f{2 \dot{M}_{\rm BH}}{M_{BH}} \left(1+ \f{Q M_{p}^{4}}{2 M_{\rm BH}^{2}}+ \f{3 Q^2 M_{p}^{8}}{8 M_{\rm BH}^{4}}\right)
\label{area_change_2}~.
\eeq
Therefore, one can write, 
\beq
\dot{A}_{\rm BH}&=&  \f{2 \dot{M}_{\rm BH}}{M_{BH}}\left(\dot{M}_{\rm photon}+ \dot{M}_{\rm atom}\right)\nonumber\\
&\times & \left(1+ \f{Q M_{p}^{4}}{2 M_{\rm BH}^{2}}+ \f{3 Q^2 M_{p}^{8}}{8 M_{\rm BH}^{4}}\right)
\label{area_change_3}
\eeq
 and recast Eq.  (\ref{area_change_3}) as follows,
\beq
\dot{A}_{\rm BH}&=&\dot{A}_{\rm photon} + \dot{A}_{\rm atom}~.
\label{area_change_4}
\eeq
Note that here,
\beq
&&\dot{A}_{\rm photon(atom)}\nonumber\\&=&  \f{2 A_{\rm BH} \,\dot{M}_{\rm photon (atom)}}{M_{BH}} \nonumber\left(1+ \f{Q M_{p}^{4}}{2 M_{\rm BH}^{2}}+ \f{3 Q^2 M_{p}^{8}}{8 M_{\rm BH}^{4}}\right)\\
&\simeq&\frac{2\dot{M}_{\text{photon(atom)}}}{M_{BH}}\frac{16\pi M_{BH}^2}{M_p^4}\biggr[1-\frac{QM_p^4}{2M_{BH}^2}-\frac{Q^2M_p^8}{16M_{BH}^4}\biggr]\nonumber\\&\times&\left[1+ \f{Q M_{p}^{4}}{2 M_{\rm BH}^{2}}+ \f{3 Q^2 M_{p}^{8}}{8 M_{\rm BH}^{4}}\right]\nonumber\\
&\simeq&\frac{32\pi\dot{M}_{\text{photon(atom)}} M_{BH}}{M_p^4}\left(1+\frac{Q^2 M_p^8}{16 M_{BH}^4}\right)~.
\label{area_photon_atom}
\eeq
Eq.  (\ref{area_change_4}) represents that the rate of change in the area of the black hole is a summation of the rate of change in the area due to the photon emission and atomic cloud near the black hole. 
At this stage we consider the change in the area of the black hole only due to the photon emission and write Eq.  (\ref{hbar_entropy_4}) as follows, 
\beq
\dot{S}_\rho&=&\,\f{8 \pi M k_B}{\hbar\,M_{p}^{2}}\left(1+\f{Q^2\,M_{p}^{8}}{16\,M^{4}}\right)\sum_{\nu}\,\hbar\,\dot{\bar{n}}_{\nu}\,\nu\nonumber\\
&=&\,\f{8 \pi M k_B}{\hbar\,M_{p}^{2}}\left(1+\f{Q^2\,M_{p}^{8}}{16\,M^{4}}\right) \dot{M}_{\rm photon} \,c^2\,,
\label{hbar_entropy_5}
\eeq
where $\sum _{\nu}\,\hbar\,\dot{\bar{n}}_{\nu}\,\nu = \dot{M}_{\rm photon} \,c^2$ represents the power transported by the emitted photons. However, in our manuscript we take $c=\hbar=1$, which cast the above equation as follows, 
\beq
\dot{S}_\rho=\,\f{8 \pi M k_B}{M_{p}^{2}}\left(1+\f{Q^2\,M_{p}^{8}}{16\,M^{4}}\right) \dot{M}_{\rm photon} 
\label{hbar_entropy_6}~.
\eeq
We replace $\dot{M}_{\rm photon}$, $A_{\rm BH}$ from Eq.  (\ref{area_photon_atom}) and Eq.  (\ref{area_change_1}) respectively in the above equation with the identification $M_{BH}=M$ and allowing upto the $\mathcal{O}(Q^2)$ we obtain, 
\beq
\dot{S}_\rho \approx\,\f{k_B\,M_p^2\dot{A}_{\rm photon}}{4}\,=\frac{k_B\dot{A}_{\rm photon}}{4G}.
\label{area_entropy_1}
\eeq
%
%
%where $\dot{A}_{\rm photon}$ is described in Eq.  (\ref{area_photon_atom}). 
Eq.  (\ref{area_entropy_1}) depicts the relation between the rate of change of HBAR entropy and the area of a BBH. 

\noindent
A new feature can be observed in this regard. Unlike the 4 dimensional RN scenario, changing the sign of $Q$, modifies the measure of the radius of the outer event horizon for a BBH (note the discussion in the subsection(\ref{BBH_1}), below Eq.   (\ref{radius_2})). If $q$ ($=Q\tilde{M}_p^2$) is positive then the radius of the outer event horizon is smaller than the Schwarzschild radius. Whereas for a negative value of $q$, $r_+$ turns out to be larger than the same. 
This implies that the two BBHs carrying equal but opposite charges and same masses possess different Bekenstein Hawking entropy. 
On contrary, under the same condition the rate of change of HBAR entropy due to the outgoing photons comes out to be identical for these two BBHs (see Eq.   (\ref{area_entropy_1})).
%
%%
%It indicates that the HBAR entropy for such BBHs remains same irrespective of the sign of the tidal charge.
%%
%
Now for two standard 4 dimensional RN black holes with equal and opposite charges and same masses the HBAR and the Bekenstein-Hawking entropy remain same. 
Thus it is plausible to assert that when two black holes with identical masses and equal but opposing charges, possess different radii of the event horizon, exhibit same HBAR entropy but different Bekenstein-Hawking entropy, can be classified as the BBHs. This feature is unique to the BBHs and can never be obtained for a standard 4 dimensional RN black hole. 
\section{Wien displacement due to the HBAR}\label{WD}
\noindent In this section, we present a comparative study of the HBAR in the background of a Schwarzschild black hole and BBH via examining the possible changes in the Wein's displacement of the wavelengths of radiation. 

\noindent For a standard Schwarzschild black hole the excitation probability of the atom can be written as follows  \cite{Fulling2}, 
\beq
P_{exc}\,(\nu)|_{\rm sc}\,d\nu \approx \f{4 \pi g^2 r_g \nu}{\o^2}\,\left(1+\f{2 \nu}{\o}\right)^{-2}\,\f{d \nu}{e^{4\pi r_g \nu} - 1}.
\nonumber\\
\label{stand_fulling_1}
\eeq
In Eq.  (\ref{stand_fulling_1}), we use $r_g=\f{2 G M}{c^2}=\,2 G M=\f{2 M}{M_{p}^{2}}$ and identify the temperature of the thermal bath as $T_{\rm sc}=\f{M_{p}^{2}}{8 \pi M}$ from the thermal distribution. 
At this stage we express Eq.  (\ref{stand_fulling_1}) in terms of the wavelength of the emitted photon ($\nu=\frac{1}{\lambda}$). Substituting $\nu=\frac{1}{\lambda}$ in the right hand side of Eq.  (\ref{stand_fulling_1}) and expressing $P_{exc}(\nu)d\nu$ as $P(\lambda)\rvert_{sc}d\lambda$, we obtain
\beq
P_{exc}\,(\lambda)|_{\rm sc}\,d\lambda =\, \f{8 \pi M g^2\,(\o\,\lambda-4) }{\lambda^4\,\o^3\,M_{p}^{2}}\,\f{d \lambda}{e^{\frac{1}{\lambda T_{\rm sc}}} - 1}
\label{stand_fulling_2}~.
\eeq
%
%
%To examine the Wien's displacement, 
We aim to find the maximum excitation probability ($P_{exc}\,(\lambda)|_{\rm sc}$) with respect to the wavelength of the radiation $(\lambda)$. It is important to note that throughout our analysis $\nu\ll \omega$ and hence $\frac{\nu}{\omega}\ll 1$. Using  $\nu=\frac{1}{\lambda}$ in the above inequality, we obtain $\omega\lambda\gg 1$. Therefore, $P_{exc}\,(\lambda)|_{\rm sc}\,d\lambda$ in the above approximation takes the form
\begin{equation}
P_{exc}\,(\lambda)|_{\rm sc}\,d\lambda \simeq\, \f{8 \pi M g^2 }{\lambda^3\,\o^2\,M_{p}^{2}}\,\f{d \lambda}{e^{\frac{1}{\lambda T_{\rm sc}}} - 1}
\end{equation}
%Thus, we examine below whether the denominator $z(\lambda)$ has a minimum value with respect to $\lambda$. 
We take the denominator $z(\lambda)=\o^2 \lambda^3 \,M_{p}^{2}\,\left(e^{1/(\lambda T_{\rm sc})}- 1\right)$.
In order to find the maximum of the transition probability,  $z(\lambda)$ must have a minimum value with respect to $\lambda$, which yields
%
%
%\beq
%&&z (\lambda)=\,\o^3 \lambda^4 \,M_{p}^{2}\,\left(e^{1/(\lambda T_{\rm sc})}- 1\right)\nonumber\\
%\implies&&\f{d z(\lambda)}{d \lambda}=\,\o^3\,M_{p}^{2}\,\left[4 \lambda^3\,\left(e^{1/(\lambda T_{\rm sc})}- 1\right) - \f{\lambda^2}{T}\,e^{1/(\lambda T_{\rm sc})}\right]\nonumber~.
%\\
%\label{min_deno_1}
%\eeq
%
%
%Minimization condition of the above equation yields, 
%
%
\beq
\f{d z(\lambda)}{d \lambda}=\,0
\implies\,1- e^{-1/(\lambda_{\rm sc}\, T_{\rm sc})}=\,\f{1}{3\,\lambda_{\rm sc}\, T_{\rm sc}}~.
\label{min_deno_2}
\eeq
\noindent We replace $\lambda = \lambda_{\rm sc}$, which implies that at $\lambda_{\rm sc}$, $z(\lambda)$ is at its minimum. This further indicates that $P_{exc}\,(\lambda)|_{\rm sc}$ will be at its maximum. 
For convenience, we take $1/(\lambda_{\rm sc}\,T_{\rm sc}) = f$ and to solve the transcendental equation as in Eq.  (\ref{min_deno_2}) we plot these two functions $\left(1- e^{-f}\right)$ and $\f{f}{3}$ which yields the intersection point of these two functions.
Thus, we obtain, $f = \f{1}{\lambda_{\rm sc}\,T_{\rm sc}} = 2.82$ and note that $\f{d^2 z(\lambda)}{d \lambda^2}|_{\lambda =\lambda_{\rm sc}}\,>\,0$,  which establishes that $P_{exc}\,(\lambda)|_{\rm sc}$ is at its maximum value for $\lambda=\lambda_{\rm sc}$.  In addition the equation $\f{1}{\lambda_{\rm sc}\,T_{\rm sc}} = 2.82={\rm constant}$ confirms the Wien's displacement law. 
\noindent We run this same analysis for the transition probability in the BBH spacetime. 
We recast the transition probability of Eq.  (\ref{prob_10}) as below, 
\beq
&P_{exc}(\nu) d\nu \approx \f{8 \pi g^{2}\nu M}{M_{p}^{2}\o^{2}}\left(1-\f{4 \nu}{\o}\right)\Bigg[1+\f{3 Q^{2} M_{p}^{8}}{16 M^4} \Bigg]\f{d \nu}{e^{\nu/T_{\rm BBH}} - 1}
\nonumber\\
&\implies  P_{exc}\,(\lambda)\,d\lambda=
\f{8\,\pi g^{2} M (4-\o \lambda)}{\lambda^4\,\o^{3}\, M_{p}^{2}}\,\Bigg[ 1+\,\f{3\,Q^{2} M_{p}^{8}}{16\,M^4} \Bigg] \f{d \lambda}{e^{1/(\lambda\, T_{\rm BBH})} -1}\,,
\nonumber
\\
\label{wien_ed_1}
\eeq
where $T_{\rm BBH} \approx \f{M_{p}^{2}}{8 \pi M}\,\left(1-\f{Q^2 M_{p}^{8}}{16 M^4}\right)=T_{\rm sc}\,\left(1-\f{Q^2 M_{p}^{8}}{16 M^4}\right)$. Performing the similar analysis as done for the Schwarzschild case, we obtain, 
$\f{1}{\lambda_{\rm BBH}\,T_{\rm BBH}} = 2.82$. This leads us to write, 
\beq
\lambda_{\rm BBH}&=&\,\f{8 \pi \,M }{2.82\, M_{p}^{2}}\,\left[1-\f{q^2\,M_{p}^{8}}{16\,\widetilde{M_{p}}^{4}\,M^4}\right]\nonumber\\
&=&\,\f{8 \pi \,\b }{2.82\, M_{p}}\,\left[1-\f{q^2}{16\,\a^{4}\,\b^4}\right]~.
\label{wien_ed_2}
\eeq
In the above equation we write the black hole and the five dimensional Planck mass parameters in terms of the four-  dimensional Planck mass as $M=\,\b M_p$, $\tilde{M}_p=\alpha M_p$. As $\tilde{M}_p \ll M_p$, $\a \ll 1$ and $\b \geq 1$.  
Similarly in case of the Schwarzschild background one obtains $\lambda_{\rm sc}$ in terms of $\b$ as follows, 
\beq
\lambda_{\rm sc}=\f{8 \pi \,\b}{2.82\,M_p}.
\label{sc_alpha_beta}
\eeq
\subsection{Analysis of the Wien's displacement pattern}\label{plot_1}
\noindent In this section, we analyse the variation of $\lambda_{\rm sc\,(BBH)}$ with respect to $\b$ and $q$,  the dimensionless tidal charge of the black hole. 
 \begin{figure}[h!]
    \centering
    \includegraphics[scale=0.35]{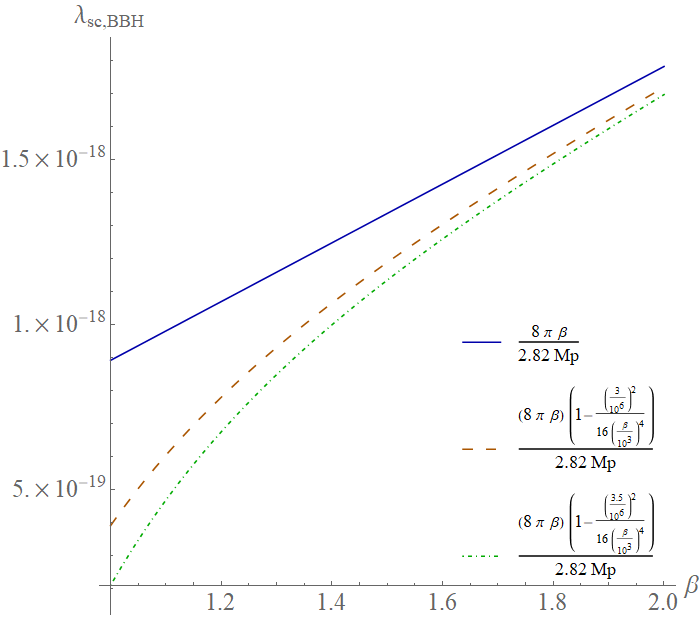}
    \caption{$\lambda_{\rm sc(BBH)}$ vs $\b$ plot for $M_p \sim 10^{19}$ GeV }
    \label{beta}
\end{figure}
 \begin{figure}[h!]
    %\centering
    \includegraphics[scale=0.35]{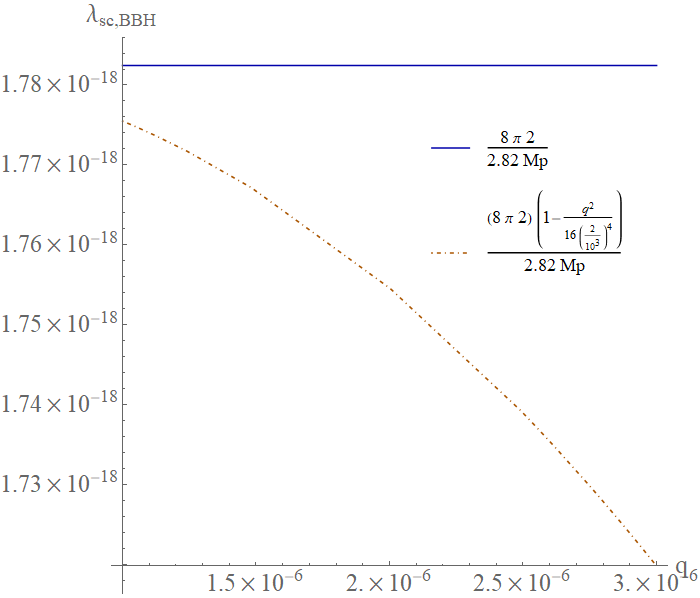}
    \caption{$\lambda_{\rm sc(BBH)}$ vs $q$ plot for $M_p \sim 10^{19}$ GeV }
    \label{q_plot}
\end{figure}
In Figure (\ref{beta}), we consider two values for $q$, such as $3.5 \times 10^{-6}$ and $3 \times 10^{-6}$,  while we keep $\a=10^{-3}$. 
The chosen values for $q$ is restricted by the conditions $q < 4 \a^2 \b^2$ from Eq.  (\ref{wien_ed_2}) and a more stringent one appears from Eq.  (\ref{radius_2}) as $q < \a^2\b^2$. Therefore, throughout this analysis we choose the parameter space for $q$ such that it complies with the latter one.  
We list our analysis as below. 
\begin{itemize}
\item
For these choices of parameters $\lambda_{\rm sc\,(BBH)}$, both become $\mathcal{O}(10^{-19})\,{\rm GeV}^{-1}$. 
Evidently, the wavelengths corresponding to the emitted photon fields for both the black hole spacetimes are extremely small and far beyond the observational capacity. 
%\textcolor{red}{Can we convert it to nm/mm to see under which EM spectra it will appear?}
%%
\item 
Although values for the $\lambda_{\rm sc\,(BBH)}$ is small, it is noteworthy that there are deviations in the values of $\lambda_{\rm sc}$ and $\lambda_{\rm BBH}$ with respect to the variation in $\b$. The deviations are more evident for the smaller values of $\b$. This suggests that for the black hole masses slightly greater or comparable to the Planck mass, the HBAR from the BBH and Schwarzschild black hole can be distinguished. 
\item
For the larger value of $\b$ the plots tentatively coincide with the standard Schwarzschild outcome. 
This dictates that for the BBHs with the masses larger than the Planck mass, the Schwarzschild potential $2M/M_{p}^{2}$ dominates over the tidal charge correction term in the metric of the BBH (see Eq.   (\ref{metric_1})) which eventually leads to the above outcome. 
\end{itemize}
Subsequently, in Figure(\ref{q_plot}), we consider $\b = 2$ and $\a = 10^{-3}$, while vary $q$ from $10^{-6}$ to $3 \times 10^{-6}$.
We list our analysis as below,
\begin{itemize}
\item
Eqs. (\ref{radius_2}) and (\ref{wien_ed_2}) dictate that 
 $q$ cannot be increased arbitrarily. Thus obeying this constraint we obtain Figure (\ref{q_plot}) which showcase an attenuation in the wavelengths with the increasing values of $q$. 
 A possible argument is that increase in the parameter $q$ implies an increment in the tidal effect, originating from the induced gravitational field on the brane. This increasing tidal effect leads to the squeezing of the wavelength of the emitted photons, which may alter the wavelengths of the emitted photons.
\item
For smaller value of $q$, the two curves corresponding to the standard Schwarzschild and BBH coincides. The Schwarzschild term will gradually become more dominating for decreasing tidal charge.
\end{itemize}
\section{Discussion}\label{discussion}
\noindent The phenomena of particle production in flat/ curved spacetime is consistently progressing since its proposition in the year 1973 \cite{Fulling21,Hawking,Hawking2,Hawking3,Unruh21,Unruh22}.
Plethora of novel outcomes, new techniques have been divulged in the studies of particle production and as well for its detection. 
%Followed by the theory of Hawking radiation several exploration have been made within the domain of GR and also the alternative theory of gravity. 
%
%
In recent times, Scully etal.\cite{Scully2003zz} have demonstrated that employing the quantum optical technique an alternative mechanism can be emanated which is based on the concept of virtual transition. This mechanism is designed using the model of cavity quantum electrodynamics and it reveals some distinctive  features of the particle production due to the acceleration radiation in flat/ curved spacetime. In regard to the curved spacetime,  for example the black hole spacetime we call this radiation process as HBAR. 
%
%
%In recent times, with the work by Scully etal.\cite{Scully2003zz}, it was well perceived that an alternative mechanism which is based on the concept of virtual transition, designed using the concept of cavity quantum electrodynamics and can be studied using the quantum optical approach may
%bring out some distinguished features of the particle production due to the acceleration radiation in flat/ curved spacetime. In regard to the curved spacetime,   for example the black hole spacetime we call this radiation process as HBAR. 
%%
HBAR phenomena is well explored within the framework of GR, whereas its fate in the context of alternative theories of gravity have been rarely investigated. Thus, in this work we study the phenomena of HBAR in the background of a BBH spacetime which emerges as the lower dimensional effective theory on the visible brane from a five dimensional gravitational theory. We find the transition probability of the atoms  and the HBAR entropy while allowing upto the quadratic power of the tidal charge. Our results depict that the transition probability depends on the quadratic power of the tidal charge and thus independent of the sign of the charge. However, its dependence on the tidal charge reflects the influence of bulk curvature effect in HBAR in the background of a higher dimensional theory. 
As mentioned in section (\ref{plot_1}), we perform a comparative study between the behaviour of the wavelengths $\lambda_{\rm sc}$ and $\lambda_{\rm BBH}$ of the radiated photon fields, which corresponds to the maximum transition probability in the standard Schwarzschild and BBH spacetimes respectively. For this analysis, we follow the theory of Wien displacement. The values of $\lambda_{\rm sc\,(BBH)}$ come out to be extremely small [see Figures (\ref{beta}) and (\ref{q_plot})] and currently far beyond the observational scope. However, Figure (\ref{beta}) dictates a theoretically alluring outcome that $\lambda_{\rm sc}$ and $\lambda_{\rm BBH}$ differ from each other for a certain mass range of the black hole, that is for, $M$ slightly greater or equal to $M_p$. 
We hope that a thorough investigation of this mass range of the black holes which can be categorised as the micro size black holes may shed some light upon the gravitational effects of the background spacetime and as well the acceleration radiation in such geometries. 
Figure (\ref{q_plot}) describes the decreasing pattern of the $\lambda_{\rm BBH}$ with increasing tidal charge of the black hole. This outcome also possesses theoretical merit such as this attenuating feature of the wavelength may arise due to the increasing tidal effect (as the tidal charge $q$ is increasing in the figure (\ref{q_plot})) which emerges due to the projection of the bulk gravitational field on the $(3+1)$-dimensional visible brane.
As a future direction, we hope to report soon the prominence of HBAR for other modified theories of gravity such as the higher curvature gravity theories.

\appendix
\begin{widetext}
\section{Some important equations}
\label{app_1}
\noindent In this Appendix, we present some important equations used in the main text. We start by considering,
\beq
y=\,1+\,\f{3\,x}{2\,\o_{dl}}~.
\eeq
Thus, we have (since $\frac{3x}{2\omega_{dl}}<1$)
\beq
y^{1/3}=\left(1+\f{3 x}{2 \o_{dl}}\right)^{1/3}\approx  1+\f{x}{2 \o_{dl}}-\f{x^2}{4 \o_{dl}^{2}}+\f{5 x^3}{24\o_{dl}^{3}}
\nonumber
\\
\label{y_power_1}
\eeq
\beq
y^{-1/3}=\left(1+\f{3 x}{2 \o_{dl}}\right)^{-1/3}\approx 1-\f{x}{2 \o_{dl}}+\f{x^2}{2 \o_{dl}^{2}}-\f{7 x^3}{12 \o_{dl}^{3}}\nonumber
\\
\label{y_power_2}
\eeq
\beq
y^{2/3}=\,\left(1+\,\f{3\,x}{2\,\o_{dl}}\right)^{2/3}\approx\, 1+\f{x}{\o_{dl}}-\,\f{x^2}{4 \,\o_{dl}^{2}}+\,\f{x^3}{6 \,\o_{dl}^{3}}
\nonumber\\
\label{y_power_3}
\eeq
\beq
&&P_{{\rm exc}}=\f{4 g^{2}_{dl}}{9} \Bigg|\bigintssss_{y=1}^{\infty} d y y^{-1/3}\Bigg[-y^{1/3}+ \f{Q_{dl}}{2}\left\{y^{1/3}-y^{-1/3}\right\}+\f{3 Q_{dl}^{2}}{8}\left\{2 y^{-1/3} - y^{-1}-y^{-1/3}\right\}\Bigg]\nonumber\\
&\times&{\rm exp}\bigg[- i \nu_{dl}\bigg\{y^{2/3}+ 2 {\rm ln}\left(y^{1/3} -1\right)+2 y^{1/3}+ \f{2 y}{3}+2 Q_{dl}{\rm ln}\left(y^{1/3}-1\right)-Q_{dl}+2 Q_{dl} y^{1/3}-\f{Q_{dl} y}{3}+ 2 Q_{dl}^{2} {\rm ln}\left(1-y^{-1/3}\right)\nonumber\\
&-&\f{Q_{dl}^{2}}{2}-\f{7 Q_{dl}^{2} y^{1/3}}{4}+\f{Q_{dl}^{2} y}{4}\bigg\}\bigg]{\rm exp}\bigg[i\o_{dl}\bigg\{- \f{2 y}{3}+ Q_{dl}\left(\f{y}{3}-y^{1/3}\right)+ Q_{dl}^{2}\left(\f{3\,y^{1/3}}{2}+\f{3\,y^{-1/3}}{4}-\f{y}{4}\right)\bigg\}\bigg]\Bigg|^2
\label{prob_4}
\eeq
\section{Integration of eq.(\ref{prob_6})}\label{int_1}
\noindent From Eq.(\ref{prob_6}), we write, 
\beq
I=\,\bigintssss_{0}^{\infty}\,d x\, \left( 1- A\,x+ B\,x^2 \right)\,(1+ i p_2\,x^2)\,x^{-2 i\,\tnudl}\,e^{-i p_1\,x}
\nonumber\\
\label{I_1}
\eeq
We take, $x=\f{x'}{p_1}$ which yields Eq (\ref{I_1}) as below. 
\beq
I&=&(p_1)^{2 i\tnudl-1}\bigintssss_{0}^{\infty}d x'  \left( 1- \f{A x'}{p_1}+ \f{B x'^2}{p_{1}^{2}} \right)\nonumber\\
&\times&(x')^{-2 i \tnudl} e^{- i x'}
+ i p_2 (p_1)^{2 i\tnudl-3}\bigintssss_{0}^{\infty} d x' \left( 1- \f{A x'}{p_1}+ \f{B x'^2}{p_{1}^{2}} \right)
(x')^{2- 2 i \tnudl} e^{- i x'}
~.
\label{I_2}
\eeq
\end{widetext}
Integrating term by term we obtain, 
Let us take, 
\begin{align}
T_1&=\,\int_{0}^{\infty} dx' \,(x')^{-2 i\,\tnudl}\,e^{- i\,x'}\nonumber\\
&= \,-\,2\,\tnudl\,e^{-\pi\,\tnudl}\,\,\Gamma[-\,2\, i \tnudl]
\label{term_1}\\
T_2&= - \f{A}{p_1}\int_{0}^{\infty} dx' (x')^{1 - 2 i\tnudl}e^{- i x'}\nonumber\\
&= - \f{2 i A}{p_1} \left\{\tnudl (1 -2 i \tnudl)\right\}e^{-\pi \tnudl}\Gamma[- 2 i \tnudl]
\label{term_2}\\
T_3&=\f{B}{p_{1}^{2}}\int_{0}^{\infty} dx' (x')^{2 - 2 i \tnudl}e^{- i x'}\nonumber\\
&= \f{4\,B}{p_{1}^{2}} \,\left\{\tnudl \,(1 - i\,\tnudl)\,(1 -\,2 i\,\tnudl)\right\}\,e^{-\pi\,\tnudl}\,\,\Gamma[-\,2 i \,\tnudl]
\label{term_3}\\
T_4&=\int_{0}^{\infty} dx' (x')^{2 - 2 i \tnudl}e^{- i x'}\nonumber\\
&=4\,\left\{\tnudl \,(1 - i\,\tnudl)\,(1 -\,2 i\,\tnudl)\right\}\,e^{-\pi\,\tnudl}\,\,\Gamma[-\,2\, i \tnudl]
\label{term_4}\\
T_5&= - \f{A}{p_1}\int_{0}^{\infty} dx' (x')^{3 - 2 i \tnudl} e^{- i x'}\nonumber\\
&=\f{4 i A}{p_1} \left\{\tnudl (1 - i \tnudl) (1 - 2 i \tnudl)(3 - 2 i\tnudl)\right\} \nonumber\\
&\times e^{-\pi \tnudl}\Gamma[- 2 i \tnudl]
\label{term_5}\\
T_6&=\f{B}{p_{1}^{2}} \int_{0}^{\infty} dx' (x')^{4 - 2 i \tnudl} e^{- i x'}\nonumber\\
&=- \f{8 B}{p_{1}^{2}}\bigg\{\tnudl (1 - i \tnudl)(1 - 2 i \tnudl) (2 - i \tnudl)\nonumber\\
&(3-2 i\tnudl)\bigg\}e^{-\pi \tnudl}\Gamma[- 2 i\tnudl]
\label{term_6}
\end{align}
Using $\f{\nu_{dl}}{\o_{dl}} \ll \,1$ and allowing all expansions upto the $\mathcal{O}(Q_{dl}^{2})$, one obtains $\f{A}{p_1}\,\sim\,A$ and $\f{B}{p_{1}^{2}}\,\sim\,B$. Thus we get, 
\beq
T_1&+&T_2+T_3=\Bigg[-\f{i \nu_{dl}}{\o_{dl}}Q_{dl} (1+Q_{dl}) - 2\nu_{dl}\bigg\{1+ Q_{dl}\nonumber\\
&&\left(1+\f{\nu_{dl}}{\o_{dl}}\right)+ Q_{dl}^{2}\left(1+ \f{2 \nu_{dl}}{\o_{dl}}\right)\bigg\}\Bigg]e^{-\pi \tnudl}\Gamma[- 2 i \tnudl]
\nonumber\\
\label{1st_term}
\eeq
\beq
&T_4&+T_5+T_6=\bigg[4\nu_{dl}\bigg\{1+Q_{dl}\left(1+\f{11 \nu_{dl}}{2\o_{dl}}\right)\nonumber\\
&&+Q^{2}_{dl}\left(1+\f{11 \nu_{dl}}{\o_{dl}}\right)\bigg\}- 8 \nu_{dl}^{3}\bigg\{1+ 3 Q_{dl} \left(1+ \f{\nu_{dl}}{3\, \o_{dl}}\right)\nonumber\\
&&+ 6 Q_{dl}^{2}\left(1+ \f{2 \nu_{dl}}{3 \o_{dl}}\right)\bigg\}
+\f{6\, i \nu_{dl}}{\o_{dl}}\, Q_{dl} \left(1+ Q_{dl}\right)\nonumber\\
&& - 12\, i \,\nu_{dl}^{2} \left\{ 1+ 2 Q_{dl} \left(1+\f{\nu_{dl}}{\o_{dl}}\right) + 3 Q_{dl}^{2} \left( 1+ \f{2 \nu_{dl}}{\o_{dl}}\right)\right\}\bigg]\nonumber\\
&&e^{-\pi\,\tnudl}\,\,\Gamma[-\,2 i \,\tnudl]
\label{2nd_term}
\eeq
Eq (\ref{I_2}) yields, 
\beq
|\,I\,|^2&=&\f{1}{p_{1}^{2}}\Bigg|\bigg\{-2 \nu_{dl}+Q_{dl} \nu_{dl}\left(\f{\nu_{dl}}{\o_{dl}} - 2\right) (1+ 2 Q_{dl})\bigg\} \nonumber\\
&&- i \Bigg\{\f{2 \nu_{dl}^{3}}{\o_{dl}} Q_{dl} (1+ 3 Q_{dl})\Bigg\} \Bigg|^{2}e^{-2 \pi \tnudl}\Big|\Gamma[- 2 i \tnudl]\Big|^2
\nonumber\\
&=&\f{1}{p_{1}^{2}}\Big| G - i H \Big|^2 e^{-2 \pi \tnudl}\Big|\Gamma[- 2 i \tnudl]\Big|^2\nonumber]\\
&=&\f{1}{p_{1}^{2}}(G^2 +H^2)e^{-2 \pi \tnudl}\Big|\Gamma[- 2 i \tnudl]\Big|^2\nonumber\\
&=&\f{4 \nu_{dl}^{2}}{p_{1}^{2}}\bigg[ 1+ 2 Q_{dl} \left(1 - \f{\nu_{dl}}{2 \o_{dl}}\right)+ 5 Q_{dl}^{2} \left(1 - \f{3 \nu_{dl}}{2 \o_{dl}}\right)\bigg]\nonumber\\
&& e^{-2 \pi \tnudl} \Big|\Gamma[- 2 i \tnudl]\Big|^2
\label{I_3}~.
\eeq
In the above equation, $G=-2 \nu_{dl}+Q_{dl} \nu_{dl}\left(\f{\nu_{dl}}{\o_{dl}} - 2\right) (1+ 2\,Q_{dl})$ and $H=\f{2\, \nu_{dl}^{3}}{\o_{dl}}\,Q_{dl} (1+ 3\,Q_{dl}) $. We also use, $\Big|\Gamma[- i \,z]\Big|^2=\,\f{\pi}{z\,{\rm Sinh}(\pi\,z)}$ in the above equation where $z= 2\,\tnudl$. 
Thus we obtain, 
\beq
|\,I\,|^2&=&\f{4 \pi \nu_{dl}^{2}}{p_{1}^{2}\tnudl}\bigg[ 1+ 2 Q_{dl} \left(1 - \f{\nu_{dl}}{2\o_{dl}}\right)+ 5 Q_{dl}^{2} \left(1 - \f{3 \nu_{dl}}{2 \o_{dl}}\right)\bigg]\nonumber\\
&&\f{1}{e^{4 \pi \tnudl} - 1}
\label{I_4}
\eeq

In Eq.  (\ref{I_4}), we explicitly use the form of $p_1$ from Eq.  (\ref{diml_2}) and obtain the transition probability as, 
\beq
P_{exc}&=& \f{4\pi g^{2}_{dl} \nu_{dl}}{\o_{dl}^{2}\left(1+\f{2 \nu_{dl}}{\o_{dl}}\right)^2}\bigg[ 1+ 2 Q_{dl} \left(1 - \f{\nu_{dl}}{2 \o_{dl}}\right)+ 5 Q_{dl}^{2}\times\nonumber\\
&& \left(1 - \f{3 \nu_{dl}}{2 \o_{dl}}\right)\bigg]\left(1+ Q_{dl} + Q^{2}_{dl}\right)^{-1}
\bigg[1-\f{Q_{dl} \nu_{dl}}{2 \o_{dl}} (1\nonumber\\
&+& 5 Q_{dl})\left(1+\f{2 \nu_{dl}}{\o_{dl}}\right)^{-1}\bigg]^{-2}\,\f{1}{e^{4 \pi\,\tnudl} - 1}~.
\label{p_exc_1}
\eeq
We further approximate the above equation while allowing upto the linear order of $\f{\nu_{dl}}{\o_{dl}}$ and quadratic order of $Q_{dl}$. This yields, 
\beq
P_{exc}&=& \f{4\,\pi g^{2}_{dl} \,\nu_{dl}}{\o_{dl}^{2}\left(1+\f{2 \nu_{dl}}{\o_{dl}}\right)^2}\bigg[ 1+ Q_{dl} + 3\, Q_{dl}^{2} \left(1 + \f{4 \,\nu_{dl}}{3\, \o_{dl}}\right)\bigg]\nonumber\\
&\times &\f{1}{e^{4 \pi\,\tnudl} - 1}~.
\label{p_exc_2}
\eeq


\begin{thebibliography}{8}
%%
\bibitem{Scully2003zz}
M. O. Scully, V. V. Kocharovsky, A. Belyanin, E. Fry and F. Capasso, \href{https://link.aps.org/doi/10.1103/PhysRevLett.91.243004}{Phys. Rev. Lett. 91 (2003) 243004}.
%
%
\bibitem{Fulling2}
M. O. Scully, S. A. Fulling, D. M. Lee, D. N. Page, W. P. Schleich and A. A. Svidzinsky, \href{https://doi.org/10.1073/pnas.1807703115}{Proc. Natl. Acad. Sci. U.S.A. 115 (2018) 8131}.
%%
\bibitem{Fulling}
A. A. Svidzinsky, S. J. Ben-Benjamin, S. A. Fulling and D. N. Page, \href{https://link.aps.org/doi/10.1103/PhysRevLett.121.071301}{Phys. Rev. Lett 121 (2018) 071301}.
%
%
\bibitem{Ordonez1}
H. E. Camblong, A. Chakraborty and C. R. Ord\'o\~nez, \href{https://link.aps.org/doi/10.1103/PhysRevD.102.085010}{Phys. Rev. D 102 (2020) 085010}.
\bibitem{Ordonez2}
A. Azizi, H. E. Camblong, A. Chakraborty, C. R. Ord\'o\~nez and M. O. Scully, \href{https://link.aps.org/doi/10.1103/PhysRevD.104.065006}{Phys. Rev. D 104 (2021) 065006}.
%
%
\bibitem{Ordonez3}
A. Azizi, H. E. Camblong, A. Chakraborty, C. R. Ord\'o\~nez and M. O. Scully, \href{https://link.aps.org/doi/10.1103/PhysRevD.104.084086}{Phys. Rev. D 104 (2021) 084086}.
%
%
\bibitem{Ordonez4}
A. Azizi, H. E. Camblong, A. Chakraborty, C. R. Ord\'o\~nez and M. O. Scully, \href{https://link.aps.org/doi/10.1103/PhysRevD.104.084085}{Phys. Rev. D 104 (2021) 084085}.
%
%
\bibitem{OTM}
S. Sen, R. Mandal and S. Gangopadhyay, \href{https://link.aps.org/doi/10.1103/PhysRevD.105.085007}{Phys. Rev. D 105 (2022) 085007}.
%
%
\bibitem{OTM2}
S. Sen, R. Mandal and S. Gangopadhyay, \href{https://link.aps.org/doi/10.1103/PhysRevD.106.025004}{Phys. Rev. D 106 (2022) 025004}.
%
%

%\cite{Das:2022qpx}
\bibitem{Das:2022qpx}
A.~Das, S.~Sen and S.~Gangopadhyay,
%``Virtual transitions in an atom-mirror system in the presence of two scalar photons,''
\href{https://link.aps.org/doi:10.1103/PhysRevD.107.025009}{Phys. Rev. D 107 (2023) 025009}.
%doi:10.1103/PhysRevD.107.025009
%[arXiv:2208.12021 [quant-ph]].
%0 citations counted in INSPIRE as of 15 Sep 2023


%\cite{Clifton:2011jh}
\bibitem{Clifton:2011jh}
T.~Clifton, P.~G.~Ferreira, A.~Padilla and C.~Skordis,
%``Modified Gravity and Cosmology,''
\href{https://doi.org/10.1016/j.physrep.2012.01.001}{Phys. Rept. 513 (2012) 1-189}.
%doi:10.1016/j.physrep.2012.01.001
%[arXiv:1106.2476 [astro-ph.CO]].
%3225 citations counted in INSPIRE as of 15 Sep 2023
%%
%\cite{SupernovaCosmologyProject:1998vns}
\bibitem{SupernovaCosmologyProject:1998vns}
S.~Perlmutter \textit{et al.} [Supernova Cosmology Project],
%``Measurements of $\Omega$ and $\Lambda$ from 42 high redshift supernovae,''
\href{https://dx.doi.org/10.1086/307221}{Astrophys. J. 517 (1999) 565}.
%doi:10.1086/307221
%[arXiv:astro-ph/9812133 [astro-ph]].
%15364 citations counted in INSPIRE as of 15 Sep 2023
%Copy to ClipboardDownload
%%
%\cite{SupernovaSearchTeam:1998fmf}
\bibitem{SupernovaSearchTeam:1998fmf}
A.~G.~Riess \textit{et al.} [Supernova Search Team],
%``Observational evidence from supernovae for an accelerating universe and a cosmological constant,''
\href{https://dx.doi.org/10.1086/300499}{Astron. J. 116 (1998) 1009.}
%doi:10.1086/300499
%[arXiv:astro-ph/9805201 [astro-ph]].
%15549 citations counted in INSPIRE as of 15 Sep 2023
%%
%\cite{Randall:1999vf}
\bibitem{Randall:1999vf}
L.~Randall and R.~Sundrum,
%``An Alternative to compactification,''
\href{https://link.aps.org/doi/10.1103/PhysRevLett.83.4690}{Phys. Rev. Lett. 83 (1999) 4690}.
%doi:10.1103/PhysRevLett.83.4690
%[arXiv:hep-th/9906064 [hep-th]].
%7451 citations counted in INSPIRE as of 15 Sep 2023
%%

%\cite{Maartens:2003tw}
\bibitem{Maartens:2003tw}
R.~Maartens,
%``Brane world gravity,''
\href{https://doi.org/10.12942/lrr-2004-7}{Living Rev. Rel. 7 (2004) 7.}
%doi:10.12942/lrr-2004-7
%[arXiv:gr-qc/0312059 [gr-qc]].
%823 citations counted in INSPIRE as of 15 Sep 2023

%\cite{Langlois:2002bb}
\bibitem{Langlois:2002bb}
D.~Langlois,
%``Brane cosmology: An Introduction,''
\href{https://doi.org/10.1143/PTPS.148.181}{Prog. Theor. Phys. Suppl. 148 (2003) 181.}
%doi:10.1143/PTPS.148.181
%[arXiv:hep-th/0209261 [hep-th]].
%336 citations counted in INSPIRE as of 15 Sep 2023

%\cite{Brax:2004xh}
\bibitem{Brax:2004xh}
P.~Brax, C.~van de Bruck and A.~C.~Davis,
%``Brane world cosmology,''
\href{https://dx.doi.org/10.1088/0034-4885/67/12/R02}{Rept. Prog. Phys. 67 (2004) 2183.}
%doi:10.1088/0034-4885/67/12/R02
%[arXiv:hep-th/0404011 [hep-th]].
%298 citations counted in INSPIRE as of 15 Sep 2023

%\cite{Rubakov:2001kp}
\bibitem{Rubakov:2001kp}
V.~A.~Rubakov,
%``Large and infinite extra dimensions: An Introduction,''
\href{https://dx.doi.org/10.1070/PU2001v044n09ABEH001000}{Phys. Usp. 44 (2001) 871.}
%doi:10.1070/PU2001v044n09ABEH001000
%[arXiv:hep-ph/0104152 [hep-ph]].
%667 citations counted in INSPIRE as of 15 Sep 2023

%\cite{Dadhich:2000am}
\bibitem{Dadhich:2000am}
N.~Dadhich, R.~Maartens, P.~Papadopoulos and V.~Rezania,
%``Black holes on the brane,''
\href{https://doi.org/10.1016/S0370-2693(00)00798-X}{Phys. Lett. B 487 (2000) 1.}
%doi:10.1016/S0370-2693(00)00798-X
%[arXiv:hep-th/0003061 [hep-th]].
%518 citations counted in INSPIRE as of 15 Sep 2023

%\cite{Emparan:2008eg}
\bibitem{Emparan:2008eg}
R.~Emparan and H.~S.~Reall,
%``Black Holes in Higher Dimensions,''
\href{https://doi.org/10.12942/lrr-2008-6}{Living Rev. Rel. 11 (2008) 6.}
%doi:10.12942/lrr-2008-6
%[arXiv:0801.3471 [hep-th]].
%677 citations counted in INSPIRE as of 15 Sep 2023

%\cite{Maartens:2010ar}
\bibitem{Maartens:2010ar}
R.~Maartens and K.~Koyama,
%``Brane-World Gravity,''
\href{https://doi.org/10.12942/lrr-2010-5}{Living Rev. Rel. 13 (2010) 5.}
%doi:10.12942/lrr-2010-5
%[arXiv:1004.3962 [hep-th]].
%559 citations counted in INSPIRE as of 15 Sep 2023
%Copy to ClipboardDownload

%\cite{Kanti:2004nr}
\bibitem{Kanti:2004nr}
P.~Kanti,
%``Black holes in theories with large extra dimensions: A Review,''
\href{https://doi.org/10.1142/S0217751X04018324}{Int. J. Mod. Phys. A 19 (2004) 4899.}
%doi:10.1142/S0217751X04018324
%[arXiv:hep-ph/0402168 [hep-ph]].
%488 citations counted in INSPIRE as of 15 Sep 2023

%\cite{Hossenfelder:2003dy}
\bibitem{Hossenfelder:2003dy}
S.~Hossenfelder, M.~Bleicher, S.~Hofmann, H.~Stoecker, and A.~V.~Kotwal,
%``Black hole relics in large extra dimensions,''
\href{https://doi.org/10.1016/S0370-2693(03)00835-9}{Phys. Lett. B 566 (2003) 233}.
%doi:10.1016/S0370-2693(03)00835-9
%[arXiv:hep-ph/0302247 [hep-ph]].
%36 citations counted in INSPIRE as of 15 Sep 2023

%\cite{Zhou:2011vy}
\bibitem{Zhou:2011vy}
K.~Zhou, Z.~Y.~Yang, D.~C.~Zou and R.~H.~Yue,
%``Spherically symmetric gravitational collapse of a dust cloud in Einstein-Gauss-Bonnet Gravity,''
\href{https://doi.org/10.1142/S0217732311036449}{Mod. Phys. Lett. A 26 (2011) 2135.}
%doi:10.1142/S0217732311036449
%[arXiv:1107.2728 [gr-qc]].
%17 citations counted in INSPIRE as of 15 Sep 2023

%\cite{Harmark:2004rm}
\bibitem{Harmark:2004rm}
T.~Harmark,
%``Stationary and axisymmetric solutions of higher-dimensional general relativity,''
\href{https://link.aps.org/doi/10.1103/PhysRevD.70.124002}{Phys. Rev. D 70 (2004) 124002.}
%doi:10.1103/PhysRevD.70.124002
%[arXiv:hep-th/0408141 [hep-th]].
%199 citations counted in INSPIRE as of 15 Sep 2023

%\cite{Shiromizu:1999wj}
\bibitem{Shiromizu:1999wj}
T.~Shiromizu, K.~I.~Maeda and M.~Sasaki,
%``The Einstein equation on the 3-brane world,''
\href{https://link.aps.org/doi/10.1103/PhysRevD.62.024012}{Phys. Rev. D 62 (2000) 024012.}
%doi:10.1103/PhysRevD.62.024012
%[arXiv:gr-qc/9910076 [gr-qc]].
%1510 citations counted in INSPIRE as of 15 Sep 2023

%\cite{Maartens:2000fg}
\bibitem{Maartens:2000fg}
R.~Maartens,
%``Cosmological dynamics on the brane,''
\href{https://link.aps.org/doi/10.1103/PhysRevD.62.084023}{Phys. Rev. D 62 (2000) 084023.}
%doi:10.1103/PhysRevD.62.084023
%[arXiv:hep-th/0004166 [hep-th]].
%413 citations counted in INSPIRE as of 15 Sep 2023

%\cite{Whisker:2008kk}
\bibitem{Whisker:2008kk}
R.~Whisker,
%``Braneworld Black Holes,''
\href{https://doi.org/10.48550/arXiv.0810.1534}{arXiv:0810.1534 [gr-qc].}
%26 citations counted in INSPIRE as of 15 Sep 2023

\bibitem{Fulling21}
S. A. Fulling, \href{https://link.aps.org/doi/10.1103/PhysRevD.7.2850}{Phys. Rev. D 7 (1973) 2850}.

\bibitem{Hawking}
S.W. Hawking, \href{https://doi.org/10.1038/248030a0}{Nature 248 (1974) 30}.
%
%
\bibitem{Hawking2}
S.W. Hawking, \href{https://doi.org/10.1007/BF02345020}{Commun. Math. Phys 43 (1975) 199}.
%
%
\bibitem{Hawking3}
S.W. Hawking, \href{https://link.aps.org/doi/10.1103/PhysRevD.13.191}{Phys. Rev. D 13 (1976) 191}.

\bibitem{Unruh21}
W. G. Unruh, \href{https://link.aps.org/doi/10.1103/PhysRevD.14.870}{Phys. Rev. D 14 (1976) 870}.
%
%
\bibitem{Unruh22}
W. G. Unruh, \href{https://link.aps.org/doi/10.1103/PhysRevD.15.365}{Phys. Rev. D 15 (1977) 365}.











\end{thebibliography}
\end{document}